\documentclass[twocolumn]{aastex631}

\hypersetup{linkcolor=blue,citecolor=blue,filecolor=green,urlcolor=magenta}

\def\mone{{\rm M}_1}
\def\mtwo{{\rm M}_2}

\def\aone{{\rm a}_1}
\def\atwo{{\rm a}_2}

\def\xeff{\chi_{\rm eff}}
\def\xeffmean{\langle\chi_{\rm eff}\rangle}
\def\xeffsd{S_X}
\def\xeffmeanp{\langle\chi_{\rm eff}\rangle^\prime}
\def\xeffsdp{S_X^\prime}
\def\xeffmeana{\langle\chi_{\rm eff}\rangle^a}
\def\xeffsda{S_X^a}
\def\fxneg{f_X-}
\def\fxnegp{f_X^\prime-}
\def\fxnega{f_X^a-}

\def\unif{{\mathcal U}}

\def\ahigh{a_{\rm high}}
\def\alow{{a_0}}
\def\thiso{\theta_{\rm i,iso}}
\def\thnk{\theta_{\rm i,nk}}
\def\pnk{P_{\rm i, nk}}
\def\thdyn{\theta_{\rm dyn}}
\def\pdyn{P_{\rm dyn}}
\def\thprim{\theta_{\rm i,prim}}
\def\pprim{P_{\rm i, prim}}
\newcommand{\Ms}{\ensuremath{{\rm M}_{\odot}}}
\newcommand{\Zs}{\ensuremath{{\rm Z}_{\odot}}}
\newcommand{\eg}{{\it e.g.}}

\newcommand{\ie}{{\it i.e.}}

\newcommand{\beq}{\begin{equation}}
\newcommand{\eeq}{\end{equation}}

\newcommand{\kmps}{\ensuremath{{\rm~km~s}^{-1}}}

\newcommand{\mcl}{\ensuremath{M_{\rm cl}}}
\newcommand{\rh}{\ensuremath{r_{\rm h}}}

\newcommand{\nbseven}{{\tt NBODY7}}

\newcommand{\bse}{{\tt BSE}}

\newcommand{\startrack}{{\tt StarTrack}}

\newcommand{\archain}{{\tt ARCHAIN}}

\newcommand{\casea}{{\tt a}}
\newcommand{\caseb}{{\tt b}}
\newcommand{\casebb}{{\tt b}$^\prime$}
\newcommand{\bone}{{\tt b1}}
\newcommand{\btwoymc}{{\tt b2ymc}}
\newcommand{\btwoiso}{{\tt b2iso}}

\newcommand{\fbin}{\ensuremath{f_{\rm bin}}}

\newcommand{\fobin}{\ensuremath{f_{\rm Obin}}}

\shorttitle{Symmetry breaking in clusters and field binaries}
\shortauthors{Banerjee, Olejak, and Belczynski}

\begin{document}

\title{Symmetry breaking in merging binary black holes
from young massive clusters and isolated binaries}

\author[0000-0002-1254-2603]{Sambaran Banerjee}
\affiliation{
Helmholtz-Instituts f\"ur Strahlen- und Kernphysik,
Nussallee 14-16, D-53115 Bonn, Germany
}
\affiliation{
Argelander-Institut f\"ur Astronomie,
Auf dem H\"ugel 71, D-53121, Bonn, Germany
}
\email{sambaran.banerjee@gmail.com}

\author[0000-0002-6105-6492]{Aleksandra Olejak}
\affiliation{
Nicolaus Copernicus Astronomical Center, Polish Academy of Sciences,
Bartycka 18, 00-716 Warsaw, Poland
}
\email{aleksandra.olejak@wp.pl}

\author{Krzysztof Belczynski}
\affiliation{
Nicolaus Copernicus Astronomical Center, Polish Academy of Sciences,
Bartycka 18, 00-716 Warsaw, Poland
}
\email{chrisbelczynski@gmail.com}

\begin{abstract}

Properties of the to-date-observed binary black hole (BBH) merger events suggest a preference towards
spin-orbit aligned mergers. Naturally, this has caused widespread interest and speculations
regarding implications on various merger formation channels.
Here we show that (i) not only the BBH-merger population from isolated binaries but also 
(ii) BBH population formed in young massive clusters (YMC) would possess an asymmetry in 
favour of aligned mergers, in the distribution of the events' effective spin parameter 
($\xeff$).
In our analysis, we utilize BBH-merger outcomes from state-of-the-art N-body evolutionary models
of YMCs and isolated binary population synthesis. We incorporate, for the first time in such an analysis,
misalignments due to both natal kicks and dynamical encounters. The YMC $\xeff$ distribution
has a mean (an anti-aligned merger fraction) of $\xeffmean\leq0.04$
($\fxneg\approx40\%$), which is smaller (larger) than but consistent with the observed asymmetry of
$\xeffmean\approx0.06$ ($\fxneg\approx28\%$) as obtained from the population
analysis by the LIGO-Virgo-KAGRA collaboration.
In contrast, isolated binaries alone tend to produce a much stronger asymmetry; for
the tested physical models,
$\xeffmean\approx0.25$ and $\fxneg\lesssim7\%$.
Although the YMC $\xeff$ distribution is more similar to the observed counterpart,
none of the channels correctly reproduce the observed distribution. 
Our results suggest that further extensive model explorations for both isolated-binary 
and dynamical channels as well as better observational constraints are necessary to 
understand the physics of `the symmetry breaking' of the BBH-merger population.

\end{abstract}

\keywords{Stellar mass black holes (1611); Massive stars (732); N-body simulations (1083);
Gravitational wave sources (677); Close binary stars (254); Young massive clusters (2049)}

\section{Introduction}\label{intro}

We are on the verge of a `golden era' of gravitational-wave (GW) and multi-messenger astronomy
\citep{Branchesi_2016,Mapelli_2018,Meszaros_2019,Mandel_2021,Spera_2022}.
Until now,
the LIGO-Virgo-KAGRA collaboration \citep[LVK;][]{Asai_2015,Acernese_2015,KAGRA_2020}
has published, in their GW transient catalogue (GWTC)
\footnote{\url{https://www.gw-openscience.org/eventapi/html/GWTC/}},
nearly 90 candidates
of general relativistic (GR) compact binary merger events. The current GWTC
includes all event candidates
from LVK's first, second (`O1', `O2'; \citealt{Abbott_GWTC1}),
and third (`O3'; \citealt{Abbott_GWTC2,Abbott_GWTC3}) observing runs,
including those in their `Deep Extended Catalogue' \citep{Abbott_GWTC2.1}.

In this study, we focus on a specific feature of GWTC, namely, the apparent
asymmetry around zero of the distribution of the effective spin parameters
in the observable merging binary black hole (BBH) population.
The effective spin parameter \citep{Ajith_2011}, $\xeff$, of a GR-merging binary
is a measure of the spin-orbit alignment of the system and is defined as
\begin{equation}
\xeff \equiv \frac{\mone|\vec\aone|\cos\theta_1 + \mtwo|\vec\atwo|\cos\theta_2}{\mone+\mtwo}
	= \frac{|\vec\aone|\cos\theta_1 +  q|\vec\atwo|\cos\theta_2}{1 + q}.
\label{eq:xdef}
\end{equation}
Here, the merging masses $\mone$, $\mtwo$, with mass ratio
$q\equiv\mtwo/\mone$ ($q\leq1$), have, respectively,
Kerr vectors $\vec\aone$, $\vec\atwo$ that project with angles
$\theta_1$, $\theta_2$ on the orbital angular momentum vector just
before the merger.

LVK's recent population analyses suggest that merging BBHs of the Universe can be
both aligned ($\xeff>0$) or anti-aligned ($\xeff<0$). However,
aligned mergers are preferred over their anti-aligned counterparts.
This is apparent from, \eg, Figure 16 of \citet{Abbott_GWTC3_prop}.
It is important to note that the observed GW events do \emph{not} support a predominantly
highly aligned or anti-aligned BBH merger population either; the $\xeff$ distribution in
\citet{Abbott_GWTC3_prop} is only slightly asymmetric around zero with a 
mean of $\xeffmean=0.06$, a standard deviation (SD) of $\xeffsd=0.10$,
and a fraction of mergers with $\xeff<0$ of 28\% (see Table~\ref{tab:distprops}; below).
Naturally, this peculiarity has sparked widespread interest 
in communities that study various mechanisms of forming merging compact binaries.

Notably, some previous analyses of GW population such as \citet{Galaudage2021} and \citet{Roulet2021},
covering events up to the LIGO–Virgo O3a observing run, find no evidence for negative-$\xeff$ systems, with the subpopulation
of $\xeff < 0$ BBH mergers being model-dependent and coming from a population of mergers involving vanishing-spin BHs.
They also find that a symmetric $\xeff<0$ distribution is rather disfavored
and that the concurrent event population can be reconstructed
by isolated-binary evolution alone. However, the latest LVK population study \citep{Abbott_GWTC3_prop},
that includes all O3 events, as well as a few other alternative analyses, \eg, \citet{Callister2022,Callister2023}
infer that BH spin orientations span a wide range of spin-orbit misalignment angles, finding evidences for significantly misaligned
and anti-aligned mergers.
In particular, \citet{Callister2022} and \citet{Abbott_GWTC3_prop} also apply an `extended' BH-spin
ansatz that explicitly includes a population of vanishing-spin BBH mergers, as in \citet{Galaudage2021},
but still infer a subpopulation of events with $\xeff<0$.
On the other hand, these studies also suggest that the $\xeff$
distribution is unlikely to be completely isotropic.
The actual negative-$\xeff$ fraction is currently uncertain,
and in this paper we adopt that from the most recent LVK population study \citep{Abbott_GWTC3_prop}
(see below).

In dynamical BBH-merger formation in, \eg, globular, open,
and nuclear clusters, one generally expects a $\xeff$
distribution that is symmetric around zero, owing to the random and
uncorrelated pairing of BHs \citep[\eg,][]{Rodriguez_2018,ArcaSedda_2021b}.
But \citet{Olejak_2021} have shown that such a `symmetry breaking'
\footnote{In this work, the term symmetry breaking refers to simply
the deviation of the mean $\xeff$ from zero. More formal symmetry-related
statistics such as skewness and moments of the $\xeff$ distribution will be considered in a follow up work.}
phenomenon
is natural for merging BBHs formed out of isolated evolution of stellar binaries,
once misalignment due to the BHs' natal kicks is taken into account.
An isolated binary evolution channel may be consistent with
LVK $\xeff$ spin distribution while assuming effective angular momentum transport in massive stars
combined with the possibility of Wolf-Rayet tidal spin-up \citep[e.g.][]{Olejak_2021,Fuller_2022,Perigois_2023}.
That allows to derive a distribution dominated by low BH spins, with an appropriate fraction of high-spinning BHs.
\citet{Tauris_2022} has additionally considered the effect of BH spin-axis tossing
in their isolated binary evolution model.

\citet{Trani_2021}, on the other hand,
have demonstrated that an LVK-like symmetry breaking can as well occur solely
due to misalignment introduced to BBHs (derived from interacting star-star binaries)
via dynamical binary-single encounters inside young
star clusters. \citet{WangYH_2021} have shown that symmetry breaking
in dynamical interactions involving BBHs (and hence in their outcomes
such as BBH mergers) is natural in co-planar dynamical systems,
such as the gas disk of an active galactic nucleus (AGN), as opposed to dynamical pairing in (near)
spherical stellar clusters.

In this work, we consider potential symmetry breaking in the BBH-merger
population originating from young massive clusters (YMC) and from isolated binaries
of the Universe.
Binary evolution naturally leads to symmetry breaking, as majority of BBH mergers form 
from stars with mostly aligned spins due to binary interactions (leading to positive $\xeff$),
with some counteracting effect introduced by natal kicks that compact objects may receive
(allowing for negative $\xeff$; \eg, \citealt{Gerosa_2017,Gerosa_2021}).
By virtue of the clusters' young age ($\lesssim100$ Myr), an
observable BBH merger population, coming from YMCs, would comprise
comparable proportions of primordially paired (\ie, where the original binary
membership is maintained; see Sec.~\ref{spinmodel})
and dynamically assembled events \citep{Belczynski_2022}. The primordially
paired mergers, due to their potential past binary-interaction phase,
would introduce a non-randomness or symmetry breaking into the population.
Here, we perform a preliminary study of the extent
of this symmetry breaking, based on a set of realistic N-body-evolutionary models
of YMCs. In our calculations, we explicitly incorporate spin-orbit misalignments due to
remnant natal kick (based on a binary population synthesis model) and
dynamical encounters (based on numerical scattering experiments).
That way, this study, for the first time, incorporates both the effects
of natal kick and dynamical encounters in estimating the $\xeff$ distribution
of a merging BBH population.

This paper is organized as follows. In Sec.~\ref{binmodel} and \ref{ymcmodel}
we summarize our isolated-binary and YMC models, respectively. We describe the
models of BH spin and BBH-merger spin-orbit alignment in Sec.~\ref{spinmodel}.
Sec.~\ref{res} describes our results. In Sec.~\ref{discuss}, we summarize our
study and identify caveats and prospects for improvement.

\section{Methods}\label{method}

\subsection{Evolutionary models of isolated-binary populations}\label{binmodel}

We use a database of properties of isolated binary BBHs generated with {\tt StarTrack}
population synthesis code \citep{Belczynski_2020}. The database has already been used and
described in \citet{Olejak_2021}.
For the tested evolutionary model (hereafter CE21), we assume standard, non-conservative
common envelope (CE) development criteria \citep{Belczynski_2008}
for which the majority of BBHs form through
CE evolution. For CE outcomes, we use the $\alpha_{\rm CE}$ formalism \citep{Webbink_1984}
with orbital energy transfer for CE ejection $\alpha_{\rm CE}=1$ and binding parameter value $\lambda$ based on \cite{Xu_2010}. We adopt a $5\%$ Bondi accretion
rate onto the BHs during CE~\citep{MacLeod_2017}. We adopt the delayed core-collapse supernova (SN)
engine ~\citep{Fryer_2012} for the final compact object mass calculations and
weak mass loss from pulsation pair-instability supernovae \citep{Belczynski_2016a}.
For BH natal kick velocities, 
we adopt a Maxwellian distribution with one-dimensional dispersion $\sigma=265\kmps$ \citep{Hobbs_2005},
which is lowered by fallback \citep{Fryer_2012}
at the compact object formation. We adopt massive O/B star wind losses as in \cite{Vink_2001},
with additional LBV winds according to prescriptions listed in Sec. 2.2 of \cite{Belczynski_2010b}.
The population synthesis assumes the cosmic star formation rate and cosmic mean metallicity evolutions of
\citet{Madau_2017}; for the metallicity evolution, a Gaussian spread of 0.5 dex ($\Zs=0.014$)
is assumed \citep{Belczynski_2020}.

In Appendix ~\ref{noce}, we additionally provide results for a second model (hereafter RLOF21) which differs from CE21 only by the
much more conservative criteria for CE development introduced in \cite{Olejak_2021a}. The revised criteria \citep[based on][]{Pavlovskii_2017}, besides having more restricted conditions for the mass ratio, take into account metallicity and the donor's radius to decide whether the system enters a CE phase. The revised criteria result in a change of the dominant formation scenario for BBH mergers which, instead of CE, consists of two stable RLOF episodes.

\subsection{Many-body evolutionary models of young massive clusters}\label{ymcmodel}

In this study, the evolutionary models of YMCs, as described in
\citet[][hereafter Ba22]{Banerjee_2022}, are utilized. All the details of
these direct-N-body-computed model star clusters are elaborated in the Ba22 paper.
Therefore, only a summary is provided below. 

In Ba22, model star clusters of initial mass
$\mcl=7.5\times10^4\Ms$ (initial number of stars $N\approx1.28\times10^5$)
and initial size (half-mass radius) $\rh=2$ pc were taken to
be representatives of YMCs. Such cluster mass and size are comparable
to those of most massive Galactic and local-group YMCs and
moderate mass `super star clusters' \citep{PortegiesZwart_2010,Krumholz_2019}.
The initial density and kinematic profiles of the clusters followed
the \citet{King_1966} model. A total of 40 models with King dimensionless potential $W_0=7$ and 9
and metallicities $Z=0.0002$, 0.001, 0.005, 0.01, and 0.02 (4 models generated
with different random seeds for each $W_0,Z$ combination) were evolved for
300 Myr (see Table A.1. of Ba22).

The initial cluster models comprised zero age main sequence
(ZAMS) stars of masses $0.08\Ms\leq m_\ast\leq150.0\Ms$ and
distributed according
to the canonical initial mass function \citep[IMF;][]{Kroupa_2001}.
The overall (initial) primordial-binary fraction was taken to be $\fbin=5$\%.
However, the initial binary fraction of
O-type stars ($m_\ast\geq16.0\Ms$), which were initially paired only among themselves,
was $\fobin(0)=100$\%, consistently with the observed high binary fraction
among O-stars in young clusters and associations
\citep[\eg,][]{Sana_2011,Moe_2017}.
The O-star binaries initially followed the observed orbital-period distribution of
\citet{Sana_2011} and a uniform mass-ratio distribution.

The model clusters were evolved with the star-by-star, direct N-body evolution code
$\nbseven$ \citep{Aarseth_2012},
which has been updated in several astrophysical aspects as detailed in \citet{Banerjee_2020,Banerjee_2020c}.
The primary `engine' for stellar and binary evolution in
$\nbseven$ is $\bse$ \citep{Hurley_2000,Hurley_2002} and that for post-Newtonian
(PN) evolution of compact binaries and higher order systems is
$\archain$ \citep{Mikkola_1999,Mikkola_2008}.
In Ba22, the `F12-rapid+B16-PPSN/PSN' \citep{Fryer_2012,Belczynski_2016a,Banerjee_2020}
remnant-mass prescription was applied.

As detailed in Ba22 and \citet{Banerjee_2021},
the above set of evolutionary YMC models was then utilized to perform a
`cluster population synthesis'. This provided estimates for present-day, intrinsic
population properties of merging BBHs (and other compact-binaries) from YMCs of the Universe.
In this population analysis, a power-law cluster birth mass function with index $\alpha=-2$
and the cosmic star formation rate evolution of \citet{Madau_2017} were adopted.
Although in Ba22 the `moderate-Z' metallicity-redshift dependence of \citet{Chruslinska_2019}
were adopted, here we repeat the cluster population synthesis exercise with 
the \citet{Madau_2017} cosmic mean metallicity evolution with a Gaussian
spread of 0.5 dex, as in \citet{Belczynski_2020}.
This revised analysis ensures a proper correspondence with the isolated-binary
evolutionary models used in this work (Sec.~\ref{binmodel}),
in terms of cosmological ingredients. All results described
in this paper correspond to this updated YMC BBH-merger population.  

\begin{figure*}
\centering
\includegraphics[width=18.0cm]{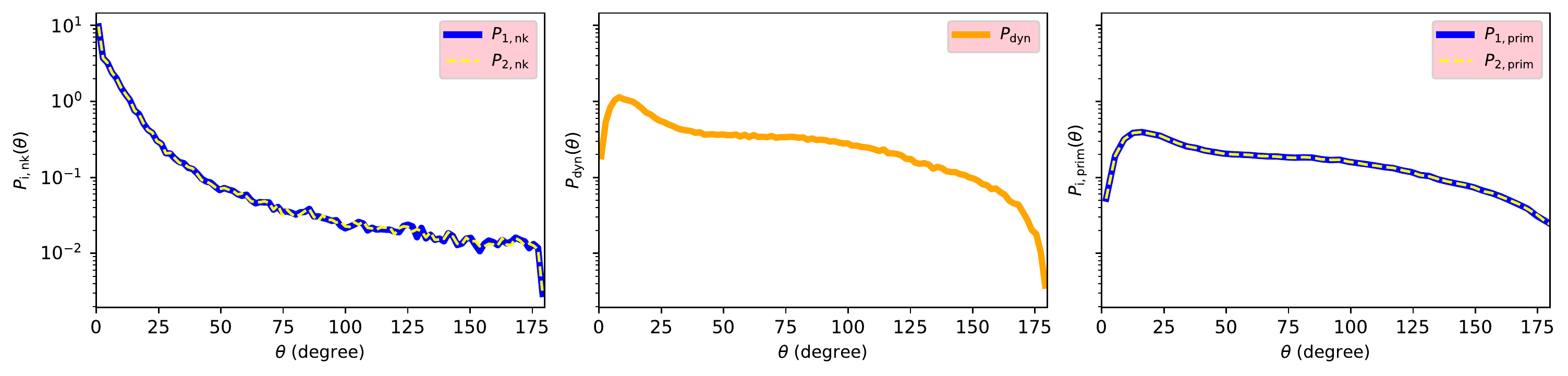}
\caption{Probability density functions for spin-orbit misalignment angles of merging BBHs, due to
	natal kicks (left panel), binary-single dynamical interactions (middle), and
	the combination of these processes (right), as considered in this work.
	These PDFs correspond to present-day, intrinsic populations of merging BBHs.}
\label{fig:theta_dist}
\end{figure*}

\subsection{Modelling spin-orbit misalignment of merging binary black holes}\label{spinmodel}

Natal BH spins in isolated binary systems are derived under the assumption of
effective angular momentum transport in massive stars, driven by the classic Tayler-Spruit dynamo~\cite {Spruit_2002}.
We adopt BH spin magnitudes being fitted to the final angular momentum of massive stellar cores
\citep[see][]{Belczynski_2020}, calculated using the MESA stellar evolutionary code ~\citep{Paxton_2015}.
Such BH natal spins (Kerr parameters) take values in the range $a \in 0.05-0.15$.
As demonstrated in \citet{Belczynski_2020}, the BH natal spins are nearly
independent of the progenitor stars' initial spins.
We allow for an efficient tidal spin-up of Wolf-Rayet (WR) stars in tight BH-WR and WR-WR binary systems,
which may significantly increase the spin, usually of the second-born BH. For such spun-up BHs,
we adopt the BH spin magnitude as in Eqn.~15 of \cite{Belczynski_2020} for the systems with an orbital period in the range
$P_{\rm orb}=0.1-1.3$ d and the maximum spin value of $a=1$ for $P_{\rm orb}<0.1$d.
As explained in \citet{Belczynski_2020}, this `spun-up' spin magnitude represents an upper limit.
BH spins may also be slightly increased due to accretion in binary systems \citep{King_2001, Mondal_2020}.
The misalignment angles of the two BH spin vectors with
respect to the orbital angular momentum is derived as described in Sec.~2.2 of \cite{Belczynski_2020}.
The initial spins of the ZAMS stars are fully aligned with their binary orbital angular momentum.
The orbit and its spatial orientation may change due to the obtained natal kick and is calculated after both BHs formation. The individual BH misalignment angles may differ due to the spin precession of binary components. We do not assume possible BH spin alignment with orbital angular momentum due to tides or mass transfer.
In this way, we expose the maximum effect of generating misalignment in BBH mergers for 
the adopted model of natal kicks.

As for the YMC models, simplistic schemes for distinguishing between non-spun-up BHs
and spun-up BHs and between BBHs depending on the formation channel
were adopted in Ba22 (see the paper for the details).
Essentially, in these models, a BH is flagged as spun-up
if, after formation, it undergoes matter accretion due to a BH-star merger or mass transfer in a binary
or if it forms in a tidally-interacting binary. In this study, we reassign the BH spins
and the BBH spin-orbit misalignment from the YMC models through post-processing
as described below.

Such cluster models with primordial binaries yield two `types' of
merging BBHs. They are (a) dynamically assembled BBHs, where the member BHs
form uncorrelated, \ie, in different primordial binaries, in a binary
formed via exchange encounter(s), or from single stars,
and (b) primordially paired BBHs\footnote{Also referred to
as `original' BBHs by some authors, \eg, \citet{DiCarlo_2020}.}, where the members
derive from the same primordial binary.
Due to the moderate
escape speed of the clusters ($\sim50\kmps$), a merged BH typically
gets ejected from such clusters right after the merger by the
associated GW recoil. No hierarchical BBH merger occurred
in these models.

The absence of hierarchical mergers, in turn, justifies post-processing of the YMC BBH-merger
populations to explore different BH-spin and alignment models.
At present, stellar-mass BHs' spin magnitudes and their distribution are far from being robustly
determined (from observed GW events or theoretical models). Therefore, inspired by
results from observed events and theoretical works, we consider several possibilities
for the YMC BHs' spins.
First, based on the bimodal BH-spin distribution as obtained from LVK's latest population
analyses \citep[][their Fig.~17]{Abbott_GWTC3_prop}, natal spins (Kerr parameters)
of the non-spun-up BHs are assigned randomly from a Maxwellian distribution peaked at $\alow=0.1$
and those of the spun-up BHs are chosen from a higher Maxwellian, peaked at $\ahigh=0.33$.
To explore uncertainties in the results, additional cases, namely, $\ahigh=0.5$ and
$\ahigh=\alow=0.1$ (no spin-up of BHs) are considered. Furthermore, we consider the
$\xeff$ distribution due to the upper limit of WR spin-up, as modelled in
our isolated binary population (the {\casebb} populations; see below).

For dynamically assembled mergers from YMCs (hereafter subpopulation {\bone}),
as in Ba22, the involved BHs are assigned isotropic and independent spin orientations, \ie,
\begin{equation}
\cos(\thiso) \in \unif(-1,1),{\rm~~~} i=1,2.
\label{eq:isotilt}
\end{equation}
Here $\unif(-1,1)$ represents a uniform probability distribution between $[-1,1]$
\footnote{To facilitate such post-processing, the BBH-merger population
from the cluster population synthesis (Sec.~\ref{ymcmodel}) are tagged,
based on information from the original N-body simulations,
to be dynamically assembled or primordially paired out of non-spun-up or spun-up
members.}.

As for primordially paired mergers (hereafter subpopulation {\btwoymc}),
the BBH would tend to be spin-orbit aligned
due to (internal) binary interactions of the parent stellar binary. However,
in general, the BBH won't be perfectly spin-orbit aligned due to
(i) natal kicks of the BHs which are generally off the binary's orbital plane (see above) and
(ii) misalignment introduced by dynamical encounters. In the YMC models,
such natal-kick and `dynamical' tilts were not explicitly considered during the N-body
computations. In this study, we model the
{\btwoymc} tilt angles as described below.

For each event in the merging BBH population, the spin-orbit misalignment
angle due to natal kick for both members are drawn independently from the
CE21 (Sec.~\ref{binmodel}) tilt angle probability distribution:
\begin{equation}
\thnk \in \pnk(\theta),{\rm~~~} i=1,2.
\label{eq:nktilt}
\end{equation}
This import is consistent, since both the isolated binary and YMC models
adopt the same fallback-modulated, momentum-conserving natal kick prescription
\citep{Belczynski_2008,Fryer_2012,Banerjee_2020}.
Dynamical encounters will introduce a common extra tilt,
\begin{equation}
\thdyn \in \pdyn(\theta),
\label{eq:dyntilt}
\end{equation}
to both BHs. We estimate $\pdyn$ from the numerical binary-single scattering experiments of
\citet{Trani_2021}, which scatterings are close passages involving mainly
resonant/chaotic interactions.
In particular, the data corresponding to their
Figure~2 is utilized to construct $\pdyn$ (averaged over metallicities)
\footnote{$\pdyn(\theta)$ represents the tilt angle distribution after a single
encounter. However, a BBH may undergo multiple encounters inside
a cluster until its in-cluster or ejected merger. Since
the encounter events are practically mutually exclusive (one encounter occurs
at a time except for very rare situations),
$\thdyn \in \sum_{k=1}^{n}\pdyn(\theta) \propto \pdyn(\theta)$.}.
Hence, the total
spin-orbit misalignment angles of the merging components of a {\btwoymc} BBH is 
(since the natal-kick and dynamical-encounter events would occur independently)
\footnote{In their original references, both $\pnk(\theta)$ and $\pdyn(\theta)$ 
are defined over $0\leq\theta\leq\pi$. In practice, $\pdyn(\theta)$ is extended
by mirror-reflecting it over $\pi\leq\theta\leq2\pi$. This effectively allows
randomly adding or subtracting $\thdyn$ to/from $\thnk$.}
\begin{equation}
\thprim = \thnk + \thdyn,{\rm~~~} i=1,2.
\label{eq:primtilt}
\end{equation}
Generally, $\thprim$ can be expected follow a probability distribution,
\begin{equation}
\thprim \in \pprim(\theta),{\rm~~~} i=1,2, 
\end{equation}
which would be shallower than both $\pnk(\theta)$ and $\pdyn(\theta)$.

\begin{figure*}
\centering
\includegraphics[width=14.0cm]{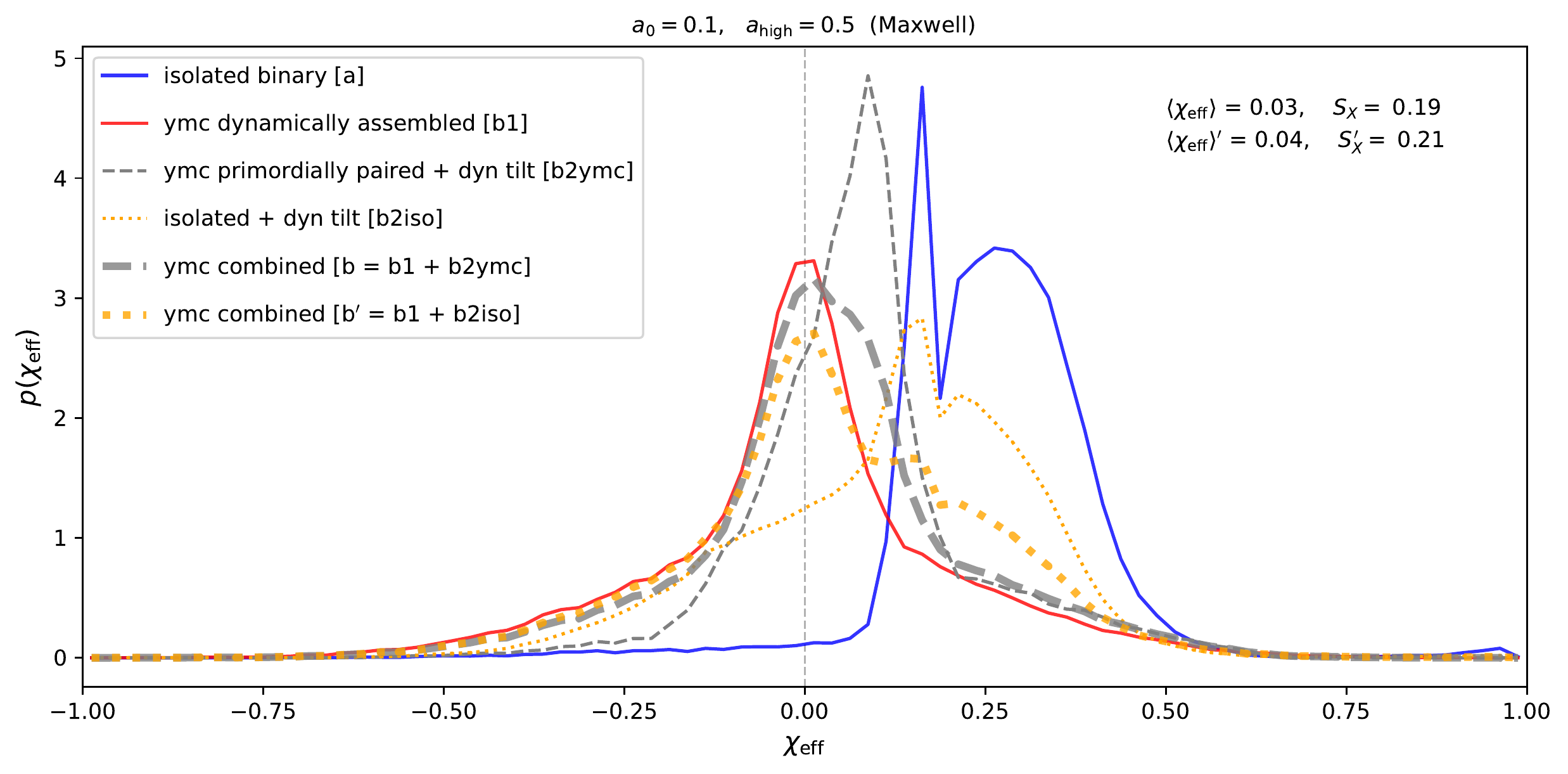}
\includegraphics[width=14.0cm]{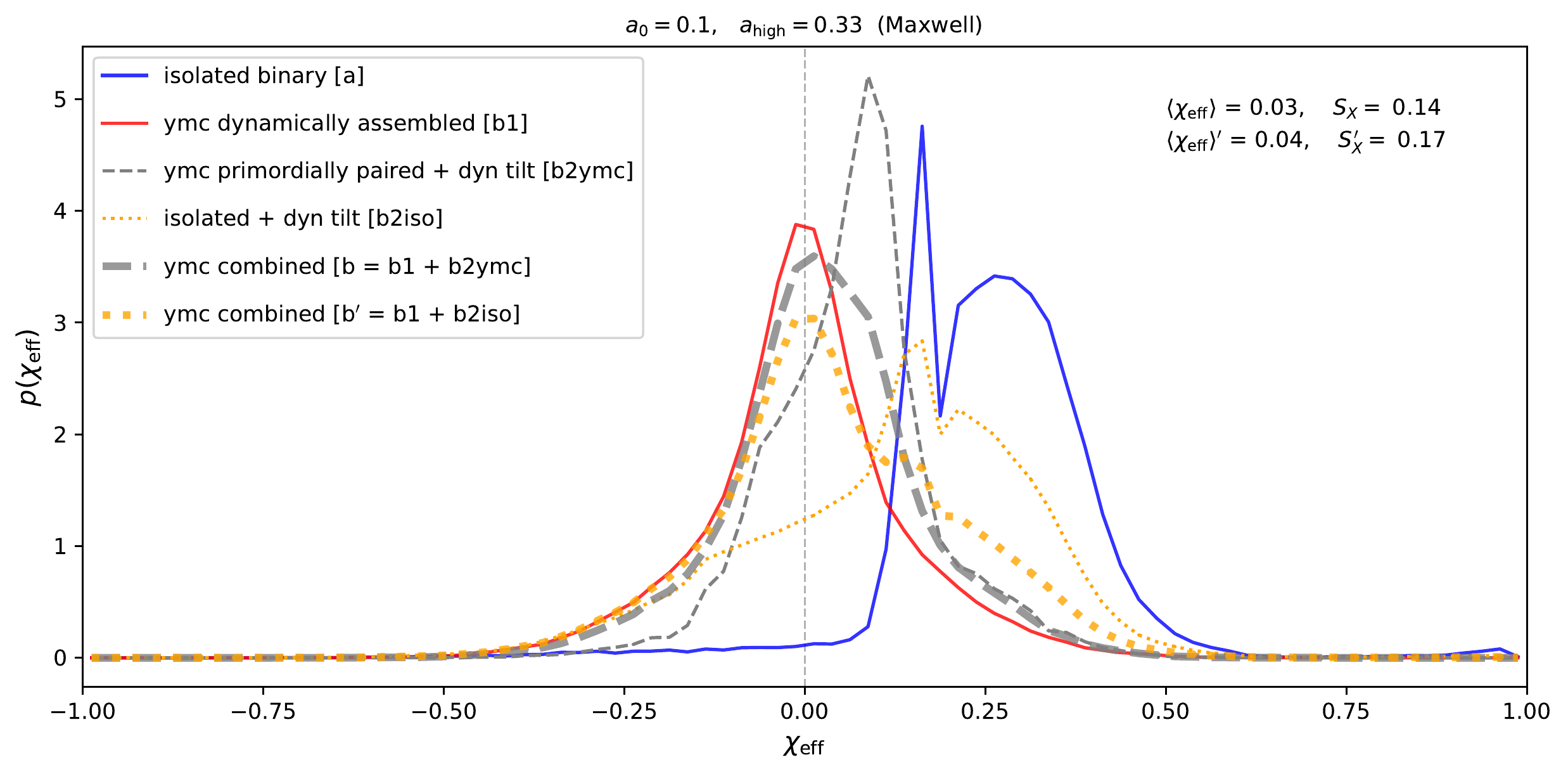}
\includegraphics[width=14.0cm]{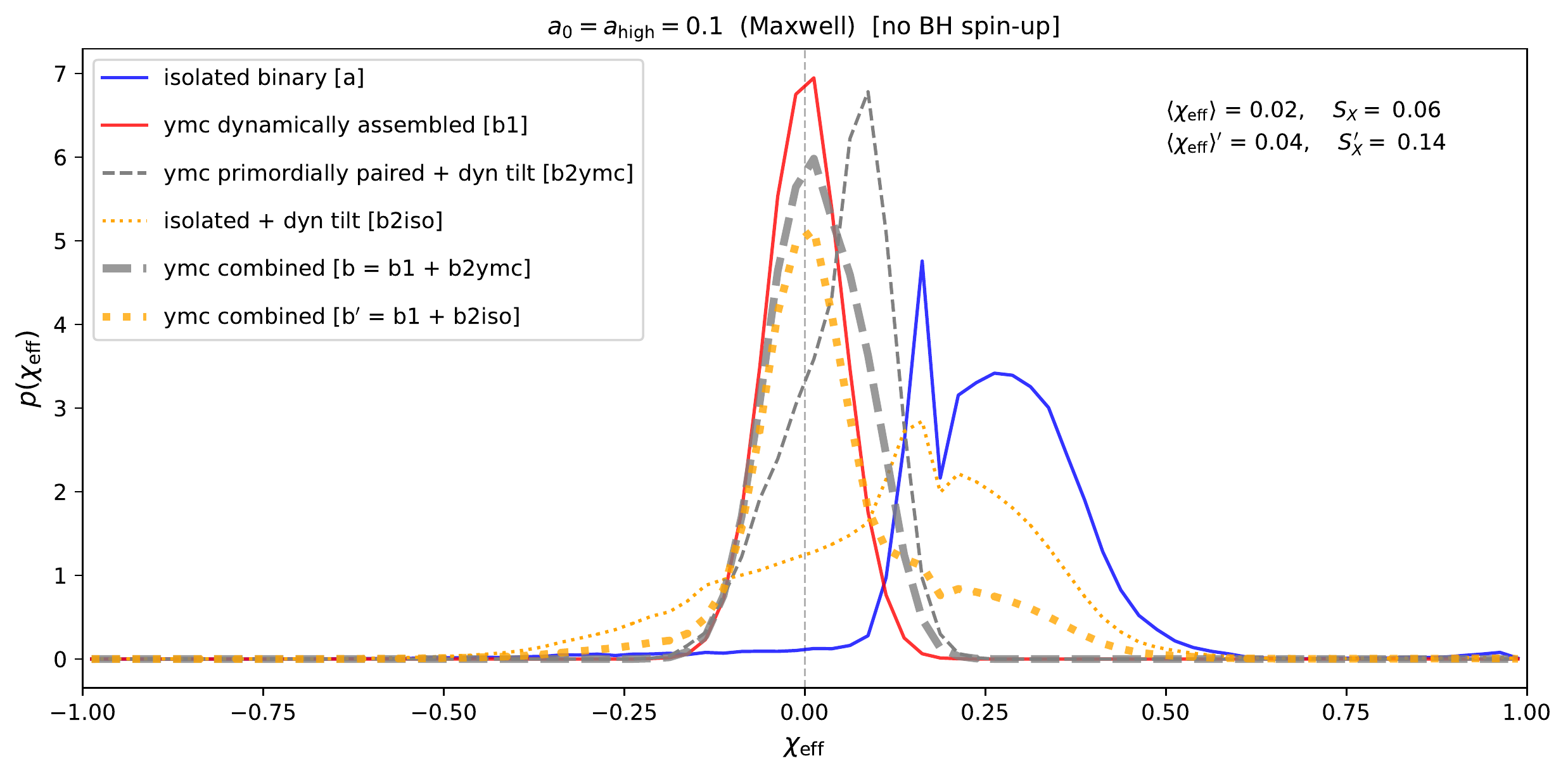}
\caption{Present-day, intrinsic distributions of $\xeff$, $p(\xeff)$, for the merging BBH populations
	from YMCs and isolated binaries. For YMCs, the $p(\xeff)$ of the dynamically assembled ({\bone}) and
	primordially paired ({\btwoymc}) subpopulations and as well as those of the combined
	populations ({\caseb} and {\casebb}) are shown separately.
	The transformation of the isolated-binary $p(\xeff)$
	due to dynamical interactions inside a cluster (\ie, of population {\casea} to
	{\btwoiso}) is also demonstrated. While plotting,
	each of the $p(\xeff)$ distribution is individually normalized to unity. The top (middle)
	panel corresponds to $\ahigh=0.5$ ($0.33$) and the bottom panel
	corresponds to the case where accretion-induced spin-up of BHs is not allowed in the YMC models.
	The population means and standard deviations of $\xeff$,
	corresponding to the combined distributions, are indicated in the upper-right corner
	of the respective panel. See text for further details.}
\label{fig:xeffdist1a}
\end{figure*}

\section{Results}\label{res}

The probability density functions (PDF) $\pnk(\theta)$, $\pdyn(\theta)$,
and $\pprim(\theta)$ are shown in Fig.~\ref{fig:theta_dist}.
The figure demonstrates that the spin-orbit alignment of merging BBHs, derived
from a population of isolated binaries, can deteriorate significantly but not completely 
if the binary population is subjected to dynamical binary-single interactions inside a cluster.

Fig.~\ref{fig:xeffdist1a} shows the corresponding $\xeff$ probability density distributions
for the {\bone} and {\btwoymc} subpopulations. The top and middle panel shows the case $\ahigh=0.5$ and
$\ahigh=0.33$, respectively. The corresponding combined distribution,
representing the overall $\xeff$ distribution of the observable merging BBH
population from YMCs (hereafter population {\caseb}), is shown.
The panels also show how the $\xeff$ distribution from
pure isolated binary evolution (\ie, from population {\casea}; see above) would transform
had these binaries been subjected to dynamical encounters inside a cluster
(hereafter subpopulation {\btwoiso}), as well as the corresponding combined
$\xeff$ distribution (hereafter population {\casebb})
\footnote{In obtaining the {\casebb} population, subpopulations {\bone} and {\btwoiso}
are mixed in the same proportions as those of {\bone} and {\btwoymc} in population {\caseb}.}.
Fig.~\ref{fig:xeffdist1a} (bottom panel)
shows the same distributions when \emph{no} spin-up of a BH due
to mass accretion or binary interaction is assumed for
the YMC models, \ie, all BHs have Kerr parameters 
drawn from a Maxwellian peaked at $\alow=\ahigh=0.1$.

The corresponding mean, $\xeffmean$ ($\xeffmeanp$), and SD, $\xeffsd$ ($\xeffsdp$), of population
{\caseb} ({\casebb}) are indicated on each panel of Fig.~\ref{fig:xeffdist1a}. 
The values of
$\xeffmean$ and $\xeffsd$ increase with increasing $\ahigh$. This is because, with increasing $\ahigh$,
the peak of the {\btwoymc} $\xeff$ distribution is shifted further towards $\xeff>0$
and, at the same time, both {\bone} and {\btwoymc} distributions become wider.
These behaviors simply follow from the definition of $\xeff$ (Eqn.~\ref{eq:xdef}) and
the assignments of the BH Kerr parameters and tilt angles for the {\bone} and
{\btwoymc} subpopulations (Sec.~\ref{spinmodel}).
The value of $\xeffmeanp$ is practically independent of $\ahigh$ since 
the asymmetry in the {\casebb} $\xeff$ distribution stems from
that of the {\btwoiso} subpopulation. Subpopulation {\btwoiso} is derived by
applying only dynamical tilts to population {\casea}
and hence is independent of the choice of $\ahigh$. The weak dependence of
$\xeffsdp$ on $\ahigh$ is due to the mixing of the (symmetric) {\bone}
subpopulation into population {\casebb}, {\bone} being $\ahigh$-dependent (Sec.~\ref{spinmodel}).

Table~\ref{tab:distprops} shows the means, SDs, and percentages of $\xeff<0$ mergers for
all the individual $\xeff$ distributions shown in Fig.~\ref{fig:xeffdist1a}. For comparison,
the corresponding LVK values are shown in the table as well
(see Secs.~\ref{intro} and \ref{discuss} for a discussion).
The LVK-observed values are obtained from the GWTC-3 public data,
utilizing the stacked $p(\xeff)$ distributions (500 of them) as read by the supplied script
{\tt make\_gaussian\_chi\_eff.py}. These $p(\xeff)$ distributions comprise
the data underlying Fig.~16 (left panel) of \citet{Abbott_GWTC3_prop} and
are obtained by them based on their `Gaussian' BH-spin population model. The set of $p(\xeff)$
corresponds to sets of $\xeffmean$, $\xeffsd$, and $\fxneg$ values.
In Table~\ref{tab:distprops}, the quoted LVK-observed values are
the unweighted mean, 5th percentile, and 95th percentile of the corresponding value-set.
(The unweighted means are very close to the corresponding
median or 50th-percentile values.)

The $\xeff$ distributions of the combined YMC BBH-merger population ({\caseb} and {\casebb}),
overall, appear similar to the observed $\xeff$ distribution. Nevertheless,
these computed populations have mean values that are somewhat smaller than the LVK's
mean $\xeff$ (Sec.~\ref{intro}) and they have negative $\xeff$ fractions
($\fxneg\approx\fxnegp\approx41\%$; see Table~\ref{tab:distprops})
somewhat higher than that of the observed BBH-merger population. Given the rather large SD values,
such small $\xeff$-asymmetries of the YMC BBH-merger population are still consistent
with the observed asymmetry.
In contrast, the $\xeff$ distribution of population {\casea} has a mean (an SD)
of $\xeffmeana=0.26$ ($\xeffsda=0.14$) and a negative-$\xeff$-merger
fraction of $\fxnega=3\%$. In other words, pure isolated binary evolution
produces a $\xeff$ distribution that is significantly more asymmetric (aligned) than the observed
$\xeff$ distribution.

The above results correspond to $\pnk$ due to the CE21 isolated binary evolution model. The RLOF21 (Sec.~\ref{binmodel}) counterparts are presented in Appendix~\ref{noce}.
As can be seen, $\pnk$ due to the RLOF21 isolated binary model leaves the results practically unaltered.
For RLOF21, $\xeffmeana=0.24$, $\xeffsda=0.18$, and $\fxnega=7\%$.
In Appendix~\ref{altspin}, we demonstrate cases with further choices
of BH spins. In particular, we limit $\alow$ and $\ahigh$ to have the same ranges
as in our isolated-binary model. As seen, the results do not alter much among
these cases.

\begin{deluxetable*}{lrrr}
\tablecaption{Statistics of the model $\xeff$ distributions.}
\tablewidth{0pt}
\tablehead{
\colhead{Merging BBH population} & \colhead{$\xeffmean$} & \colhead{$\xeffsd$} & \colhead{$\fxneg$}
}
\startdata
	LVK observed & $0.06^{+0.03}_{-0.04}$ & $0.11^{+0.04}_{-0.03}$ & $28.32^{+14.46}_{-13.21}$ \\
\hline
	isolated  binary [{\casea}] & 0.26 & 0.14 & 2.83 \\
	isolated + dyn. tilt [{\btwoiso}] & 0.12 & 0.19 & 25.32 \\
\hline
	\multicolumn{4}{c}{$\alow=0.1,{\rm ~~~~}\ahigh=0.5${\rm~~~~}(Maxwell)}\\
\hline
	ymc combined [{\caseb} = {\bone} + {\btwoymc}] & 0.03 & 0.19 & 40.69  \\
	ymc combined [{\casebb} = {\bone} + {\btwoiso}] & 0.04 & 0.21 & 40.74  \\
%	isolated [{\casea}] & 0.26 & 0.14 & 2.83 \\
%	isolated + dyn. tilt [{\btwoiso}] & 0.12 & 0.19 & 25.38 \\
	ymc dyn. assembled [{\bone}] & 0.00 & 0.20 & 50.19 \\ 
	ymc prim. paired + dyn. tilt [{\btwoymc}] & 0.08 & 0.15 & 25.10 \\
\hline
\multicolumn{4}{c}{$\alow=0.1,{\rm ~~~~}\ahigh=0.33${\rm~~~~}(Maxwell)}\\
\hline
	ymc combined [{\caseb} = {\bone} + {\btwoymc}] & 0.03  & 0.14 & 40.80  \\
	ymc combined [{\casebb} = {\bone} + {\btwoiso}] & 0.04  & 0.17 & 40.87  \\
%	isolated [{\casea}] & 0.26 & 0.14 & 2.83 \\
%	isolated + dyn. tilt [{\btwoiso}] & 0.12 & 0.19 & 25.30 \\
	ymc dyn. assembled [{\bone}] & 0.00 & 0.14 & 50.38 \\ 
	ymc prim. paired + dyn. tilt [{\btwoymc}] & 0.07 & 0.11 & 25.43 \\
\hline
	\multicolumn{4}{c}{$\alow=\ahigh=0.1${\rm~~~~}(Maxwell){\rm~~~~}[no BH spin-up]}\\
\hline
	ymc combined [{\caseb} = {\bone} + {\btwoymc}] & 0.02  & 0.06 & 40.60  \\
	ymc combined [{\casebb} = {\bone} + {\btwoiso}] & 0.04  & 0.14 & 40.66  \\
%	isolated [{\casea}] & 0.26 & 0.14 & 2.83 \\
%	isolated + dyn. tilt [{\btwoiso}] & 0.12 & 0.19 & 25.32 \\
	ymc dyn. assembled [{\bone}] & 0.00 & 0.05 & 49.95 \\ 
	ymc prim. paired + dyn. tilt [{\btwoymc}] & 0.05 & 0.07 & 24.78 \\
\enddata
\tablecomments{The columns from left to right are, respectively, BBH-merger population
	type, mean of the population's $\xeff$ distribution, standard deviation of the
	$\xeff$ distribution, and percentage of events in the population with $\xeff<0$.
	The LVK-observed values (see text)
	are the mean values along with the 90\% confidence limits.
	The YMC-population values for different $\ahigh$ (Sec.~\ref{spinmodel}; Fig.~\ref{fig:xeffdist1a})
	are shown in separate sections, as indicated.}
\label{tab:distprops}
\end{deluxetable*}

\section{Discussions and outlook}\label{discuss}

In this study, we consider the $\xeff$ distributions of observable merging BBH populations
produced by YMCs and isolated, massive stellar binaries of the Universe.
Such model populations are obtained based on population synthesis of massive binaries
(using {\startrack}; Sec.~\ref{binmodel}) and direct N-body evolutionary models of YMCs
(using {\nbseven}; Sec.~\ref{ymcmodel}). In our analysis,
we take into account spin-orbit misalignments of the merging BBHs due to
both natal kicks and dynamical encounters inside clusters (Sec.~\ref{spinmodel}).
We show that (Sec.~\ref{res}) (i) a primordially paired merging BBH population from the YMC models
introduces an asymmetry in the $\xeff$ distribution towards $\xeff>0$, (ii) the overall
bias of the YMC BBH-merger population towards aligned mergers ($\xeffmean\leq0.05$, $\fxneg\approx40\%$)
is somewhat smaller than but still comparable to that
in the observed BBH-merger population, and (iii) the $\xeff$-asymmetry in BBH mergers from isolated binaries
alone ($\xeffmeana\approx0.2$, $\fxnega\leq7\%$) is rather high compared to that in the observed population.

Among stellar clusters, young clusters such as YMCs would have the highest contribution  
towards primordially paired GR mergers, since such mergers have short delay times ($\lesssim 100$ Myr).
With time, dynamically paired mergers become increasingly dominant in clusters
(see \citealt{Belczynski_2022} and references therein).
Hence, old clusters (\eg, old open clusters, globular clusters, nuclear clusters) are
unlikely to cause an alignment bias in BBH mergers.
The above results and considerations suggest that a combination of cluster
and isolated-binary merger channels across cosmic time might yield the
observed $\xeff$ asymmetry. This possibility can be explored by a straightforward extension
of the approach described in \citet{Banerjee_2021b}, which will be taken up
in future work. The method would as well allow incorporating other
promising channels for orientation bias in BBH mergers such as
dynamics in AGN gas disks \citep{WangYH_2021,Rozner_2022}. 

The extent of the contribution of young clusters to the observed BBH-merger
the population is, at present, far from being settled. This stems from
various astrophysical and as well GW-observational uncertainties; see, \eg,
\citet{Antonini_2020b,Banerjee_2020d,Banerjee_2021,Banerjee_2021b,Chatto_2022}. The same is
true for other dynamical channels \citep[\eg,][]{Chatterjee_2017a,ArcaSedda_2020b}
and the isolated-binary channel \citep[\eg,][]{vanSon_2022,Belczynski_2022a}.
With further
GW events detected during the forthcoming observing runs, constraints on the different channels
can be expected to improve.

In this work, we present the results of one physical model (CE21) for the isolated binary evolution channel, where the vast majority of BBH mergers (over $90 \%$ of them) form via CE evolution scenario. In Appendix \ref{noce}, we provide additional results for the second, RLOF21 model, which corresponds to much more restrictive CE development criteria than the ones applied for the CE21 model. The revised criteria, motivated by the studies of \cite{Pavlovskii_2017}, change the dominant formation scenario for the BBH mergers which, instead of CE, now consists of a stable mass transfer phase ($\sim 90\%$ of systems) during the second RLOF. This modification also significantly reduces BBH merger rates at high redshifts ($z>1.0$), since the produced BBH systems tend to have wider orbital separations (also longer delay times) than those in the CE scenario, see, \eg, Figure B1 in \cite{Olejak_2022}.

Unexpectedly, the two tested models, which might be considered as extremes of the CE development criteria, result in similar distributions of the $\xeff$. A significant fraction of high-spinning BBH mergers in model RLOF21 originates from a mass-reversal scenario, that includes two episodes of stable mass transfer \cite{Olejak_2021}. However, note that similar studies, which apply CE development criteria with restrictions somewhere between our CE21 and RLOF21 models, may lead to other findings and conclusions. For example, \cite{Zevin_2022} and \cite{Bavera_2022} find that the stable mass transfer formation scenario tends to produce lower $\xeff$s than in CE evolution. The discrepancy between our findings and theirs is caused by the absence of our mass-reversal evolutionary scenario in their simulations. Instead, this scenario is common in our simulations due to the restrictive CE development criteria, allowing highly unequal mass systems (with the donor to BH mass ratio $\gtrapprox 6$) to avoid CE phase evolution \cite{Olejak_2021}. Note that, e.g., \cite{Broekgaarden2022} also proposed a mass-reversal scenario (via stable mass transfer) as a possible formation channel for a BBH merger with a non-negligible $\xeff$ value. See Appendix \ref{noce} for more details.

Our studies test only two models for isolated binary evolution channel which result in a similar distribution of $\xeff$, but a significant difference in the distribution of individual BH spin magnitudes, see Appendix \ref{noce} and \cite{Olejak_2021} for details. Besides criteria for CE development, there are many other unconstrained physical processes and parameters in rapid population synthesis codes. Some broader parameter studies, testing how different assumptions on, \eg, angular momentum transport, increased accretion rate, or CE efficiency in parameterized $\alpha_{\rm CE}$ formalism impact the fraction of high-spinning BBH mergers in isolated binary evolution in CE and stable mass transfer scenarios, may be found in \cite{vonSon_2020, Zevin_2022, Broekgaarden2022} and \cite{Perigois_2023}. A few other examples of uncertainties that may affect effective spins of BBH mergers in isolated binary evolution but has not been examined within this study are: metallicity-specific star formation rate density \citep{Santoliquido2021, Briel_2021, Chruslinska_2022}, stellar winds \citep{Vink_2001, Sander_2022}, core-collapse SN mechanism \citep{Sukhbold_2016, Fryer_2022, Olejak_2022, Shao_2022} and BH natal kicks \citep{Mandel_2021a}.

In fact, it may even be possible to reproduce the observed $\xeff$ distribution by the isolated binary evolution channel alone. Especially, our calculations of WR tidal spin-up should be considered as an upper limit  \citep{Belczynski_2020, Ma_2023} that may lead to a significant overestimation of the fraction of high-$\xeff$ BBH mergers. 
Assumptions about poorly understood BH natal kicks affect the spin-orbital misalignment, and thus the fraction of BBHs with negative $\xeff$ -- larger kicks would produce a larger fraction of negative-$\xeff$ mergers
from binary evolution, \eg, \citet{Belczynski_2020}. Also, the dynamo of \citet{Fuller_2019a}, rather than the presently used Tayler-Spruit dynamo, along with less efficient tidal spin-up of WR stars \cite{Ma_2023} would move the peak of the isolated-binary $\xeff$ distribution towards a smaller value.

It is worth noting that our isolated-binary evolutionary model does not incorporate chemically homogeneous
binary evolution. Adequately tight and massive binaries can undergo chemically homogeneous evolution,
which would result in highly spinning, near equal mass, fully aligned BBH mergers and hence contribute
towards enhancing the positive $\xeff$ bias in isolated-binary BBH mergers.
The chemically homogeneous evolution scenario \citep{deMink_2009,deMink_2010, Marchant_2016, DeMink_2016, Mandel_2016, Riley_2021} applies to initially tight, almost contact, massive binary systems with their periods below $P<2$ days \citep{DeMink_2016}. Main sequence components get tidally spun up, which induces efficient mixing from the interior throughout their envelope, providing the core with additional nuclear-burning elements. Importantly, the rapid rotation of stars and the efficient mixing is expected to result in a significant decrease in radius expansion in contrast to the non-rotating evolutionary models \citep{Maeder_1987}. That allows components to avoid the stellar merger after the completion of hydrogen burning and survive as a tight binary system \citep{Nelson_2001}. The chemically homogeneous formation scenario favors massive BBH mergers of the total mass of $M_{\rm tot} \in 50-110$ $M_{\odot}$ with secondary to primary BH mass ratio of $q \gtrsim 0.8$ \citep{Marchant_2016, DeMink_2016}.

In this work, we focus only on the $\xeff$ parameter distribution of the detected BBH mergers.
However, LVK also provides distributions of other compact-object parameters, \eg, masses, mass ratio, and as well reports
potential correlations between mass ratio and $\xeff$ of BBH mergers \citep{Abbott_GWTC3_prop}. Further studies need to be done
to reconstruct all properties of the GW-source population.

The present YMC model set from Ba22 (Sec.~\ref{ymcmodel}) has only one representative initial cluster mass and size
(although it spans over a wide range of metallicity), which is a limitation of the model grid.
A grid spanning over cluster mass and size is being computed which will eventually be
incorporated. Although the cluster model grid of \citet{Banerjee_2020c} spans over cluster
mass, size, and metallicity, the model grid of Ba22 is preferred in the present analysis
due to the latter grid's homogeneity in properties. This ensures that no asymmetry
occurs in the model BBH population due to artefacts of the cluster model grid.

Exploration of cluster parameters is important for an in-depth understanding
of the production of primordially paired mergers in clusters. The relative numbers
of primordially paired and dynamically assembled mergers directly influence the results
of this work -- a higher relative number of dynamical mergers would cause a more symmetric
$\xeff$ distribution. Therefore, the initial fraction of primordial binaries among the
BH-progenitor stars assumed in YMC models (which is 100\% in the present models;
see Sec.~\ref{ymcmodel}) would influence the resulting $\xeff$ distribution.
Cluster (initial) density can play a more subtle role: on one hand, a denser cluster
can more efficiently dynamically harden wide primordially paired
(as well as dynamically assembled) BBHs and aid them
in merging. On the other hand, higher density can also aid primordial binaries
to undergo premature star-star mergers (via dynamical hardening or hierarchy formation),
reducing the number of primordial BBH mergers. Initial orbital period distribution
of the primordial binaries can similarly influence the primordially paired BBH
merger fraction. In fact, a higher primordial binary fraction can trigger
more frequent stellar-phase mergers of primordial binaries via
binary-binary interactions that form hierarchical subsystems \citep[\eg,][]{Geller_2013}, that way
potentially weaken the dependence on primordial binary fraction.
At present, how cluster parameters and stellar content
influence the $\xeff$ distribution of BBH mergers from clusters is hardly
explored and remains mainly an open question.

The present
analysis does not identify instances where the dynamical assembly of the binary happens
before the formation of the member BHs and hence, due to a potential interacting-binary phase,
preferential alignment can be expected. This would boost the asymmetry in the YMC BBH mergers.
This effect is unlikely to be critical for the presently considered pc-scale, massive clusters
(although it will be modelled in future studies) where only the most massive members can
mass-segregate over the first $\lesssim10$ Myr. However, early dynamical pairing among massive
stars (via exchange and/or three-body encounters) is much more efficient in compact, low-mass clusters
(of $\sim0.1$ pc, $10^2 - 10^3 \Ms$ initially) with short relaxation/mass-segregation time
\citep[\eg,][]{DiCarlo_2020,Kumamoto_2020,Rastello_2021}.
Another limitation of the present analysis is that tilt due to
binary-binary (and higher-order) encounters is not incorporated. Such encounters, although are generally less
frequent than binary-single interactions, can potentially cause larger tilts
(due to longer chaotic phase and more diverse outcomes) and hence enhance the
randomization of the primordially paired BBHs' orientations.

The comparison with the observed BBH-merger population, especially of negative $\xeff$ fraction ($\fxneg$), should be taken with caution, given the former, alternative interpretation that the observed population is practically devoid of negative-$\xeff$ events \citep{Roulet2021,Galaudage2021}. Notably, such interpretation, performed for the previous GW events catalog (not including the full O3 run), relies on the assumption of a significant contribution from a zero-spin BBH merger subpopulation. This assumption rests essentially on a specific, highly efficient (driven by modified Tayler-Spruit dynamo) stellar angular momentum transport model \citep{Fuller_2019a}. A more recent, parameter-free analysis finds the distribution of BH spin magnitudes unimodal and concentrated at small but non-zero values \citep{Callister2023}. Such distribution is more in line with the stellar models that apply the classic Tayler-Spruit dynamo \citep{Spruit_2002,Belczynski_2020}.
Hence, the interpretation by LVK leading to a finite $\fxneg$
without a zero-spin-merger `spike' in the population (see Secs.~\ref{intro}, \ref{res}; Table~\ref{tab:distprops}),
as adopted here, is, potentially, more general and model-inclusive.

Neither the isolated-binary nor the YMC models in the present study properly reproduces the
observed $\xeff$ distribution of BBH mergers. However, the YMC $\xeff$ distribution is reasonably
similar to the observed $\xeff$ distribution, taking into account the observational uncertainties.
Given the observational and model uncertainties, at present it cannot be conclusively
said whether one channel or a combination of channels causes the
observed $\xeff$ distribution. This can expected to be better settled only with further
extensive model explorations and GW event detections.

\begin{acknowledgments}
We thank Alessandro A. Trani for providing the data corresponding
to Figure 2 of their Trani et al. (2021) paper.
Special thanks goes to tens of thousands of
citizen-science project "Universe@home" (universeathome.pl) enthusiasts
that help to develop the StarTrack population synthesis code used in this study.
S.B. acknowledges funding for this work
by the Deutsche Forschungsgemeinschaft (DFG, German Research Foundation)
through the project ``The dynamics of stellar-mass black holes in dense stellar systems and their
role in gravitational-wave generation'' (project number 405620641; PI: S. Banerjee).
S.B. acknowledges the generous support and efficient system maintenance of the computing teams at the
AIfA and HISKP.
AO and KB acknowledge support from the Polish National Science Center (NCN) grant
Maestro (2018/30/A/ST9/00050). AO is also supported by the Foundation for Polish Science (FNP)
and a scholarship of the Minister of Education and Science (Poland).
\end{acknowledgments}

\software{
StarTrack \citep{Belczynski_2008},
NBODY7 \citep{Aarseth_2012},
NumPy \citep{harris2020array},
matplotlib \citep{Hunter_2007}
}

\appendix

\section{The case of stable-RLOF-dominated merging BBH formation from isolated binary evolution} \label{noce}

The distributions of effective spin parameter derived for both tested CE development criteria, the less (CE21) and the more (RLOF21) restrictive, look similar. However, there is an important difference in the high-spinning BBH subpopulation between the CE and stable mass transfer formation scenarios \cite{Olejak_2021}. Once we adopt the less restrictive criteria (CE21), for which BBH mergers are formed mainly via CE evolution, the second-born and tidally spun-up BH is usually the less massive one. It is also possible that systems with almost equal-mass components pass through a close WR-WR phase, during which both components are spun up; see Figure 2 of \citet{Olejak_2021}. 

Conversely, for the restrictive CE development criteria (RLOF21), the vast majority of BBH mergers form through stable mass transfer, and the second-born tidally spun-up BH is usually the more massive one. This result is common for this formation channel of highly-spinning BBH mergers via the mass-reversal scenario; see Figure 1 of \cite{Olejak_2021}. Stable mass transfer, widely adopted by rapid population synthesis codes, is much less efficient in tightening the orbit than CE evolution. Therefore, our simulations require highly unequal mass ratio components at the onset of the second RLOF to eject enough mass and angular momentum for shortening the binary orbital period below $P \leq 1.3$ day and entering the WR tidal spin-up regime \citep{Belczynski_2020}. This also results in the characteristic mass ratio distribution of BBH mergers formed via stable mass transfer channel \citep{Olejak_2021a} which possesses a high, extensive peak between $q \approx 0.4-0.7$. 

For this reason, the formation of a significant fraction of close BH-WR star systems via stable mass transfer is possible only while adopting very restrictive CE development criteria, which allows highly unequal mass systems, such with the donor to BH mass ratio $\frac{M_{\rm don}}{M_{\rm BH}} \gtrapprox 6$, to go through stable mass transfer phase instead of CE. As our restrictive criteria are rather extreme among rapid population synthesis codes, such a mass-reversal formation scenario for high-spinning BBH mergers is unique for {\tt StarTrack} code. However, see also the results of \cite{Broekgaarden2022} who also find the mass-reversal scenario as a possible origin of BBH mergers with non-negligible $\xeff$.

Note that e.g., \cite{Zevin_2022,Bavera_2022} came to different conclusions, suggesting that once adopting effective angular momentum transport and Eddington-limited accretion, the fraction of high-spin BHs originating from a stable mass transfer channel should be smaller than for CE scenario due to the less efficient orbital tightening. In our {\tt StarTrack} simulations, the stable mass transfer scenario also tends to produce wider systems than evolution via CE, resulting in larger time delays and decreased BBH merger rates at higher redshifts, see e.g., Figure B1 in \cite{Olejak_2022}. The inconsistency between our results and those of other groups likely originates from the difference in the adopted CE development combined with a conservative assumption on angular momentum loss \citep{Olejak_2021}. Less restrictive criteria would eliminate the main formation scenario for high-spinning BHs via stable mass transfer found in our simulations, as highly unequal binaries would develop CE instead.  

\begin{figure*}
\centering
\includegraphics[width=18.0cm]{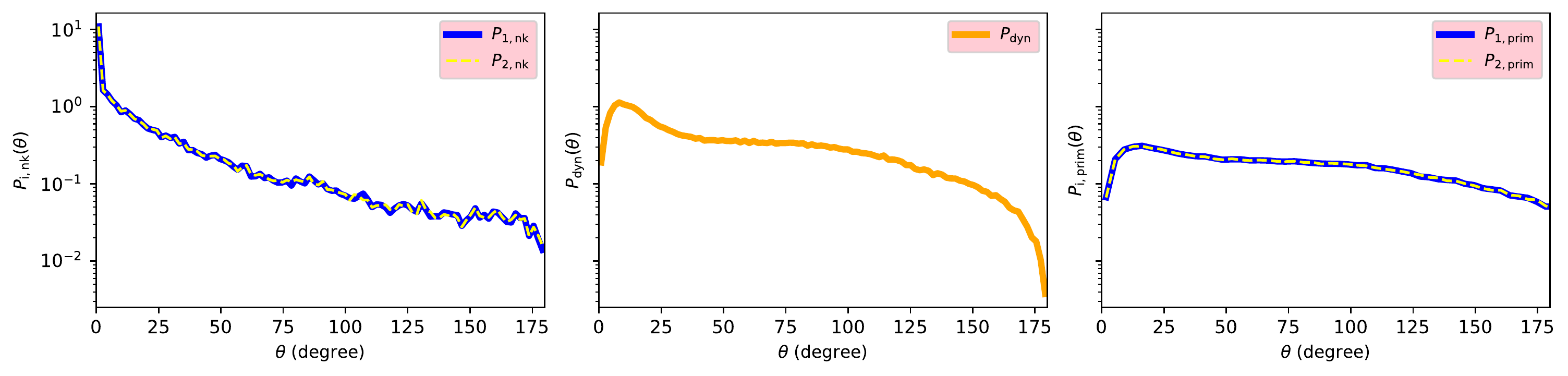}
\caption{Same description as in Fig.~\ref{fig:theta_dist} applies, except that the model
	isolated-binary population evolution forms BBH mergers primarily via
 	stable RLOF (the RLOF21 model).}
\label{fig:theta_dist_noce}
\end{figure*}

\begin{figure*}
\centering
\includegraphics[width=14.0cm]{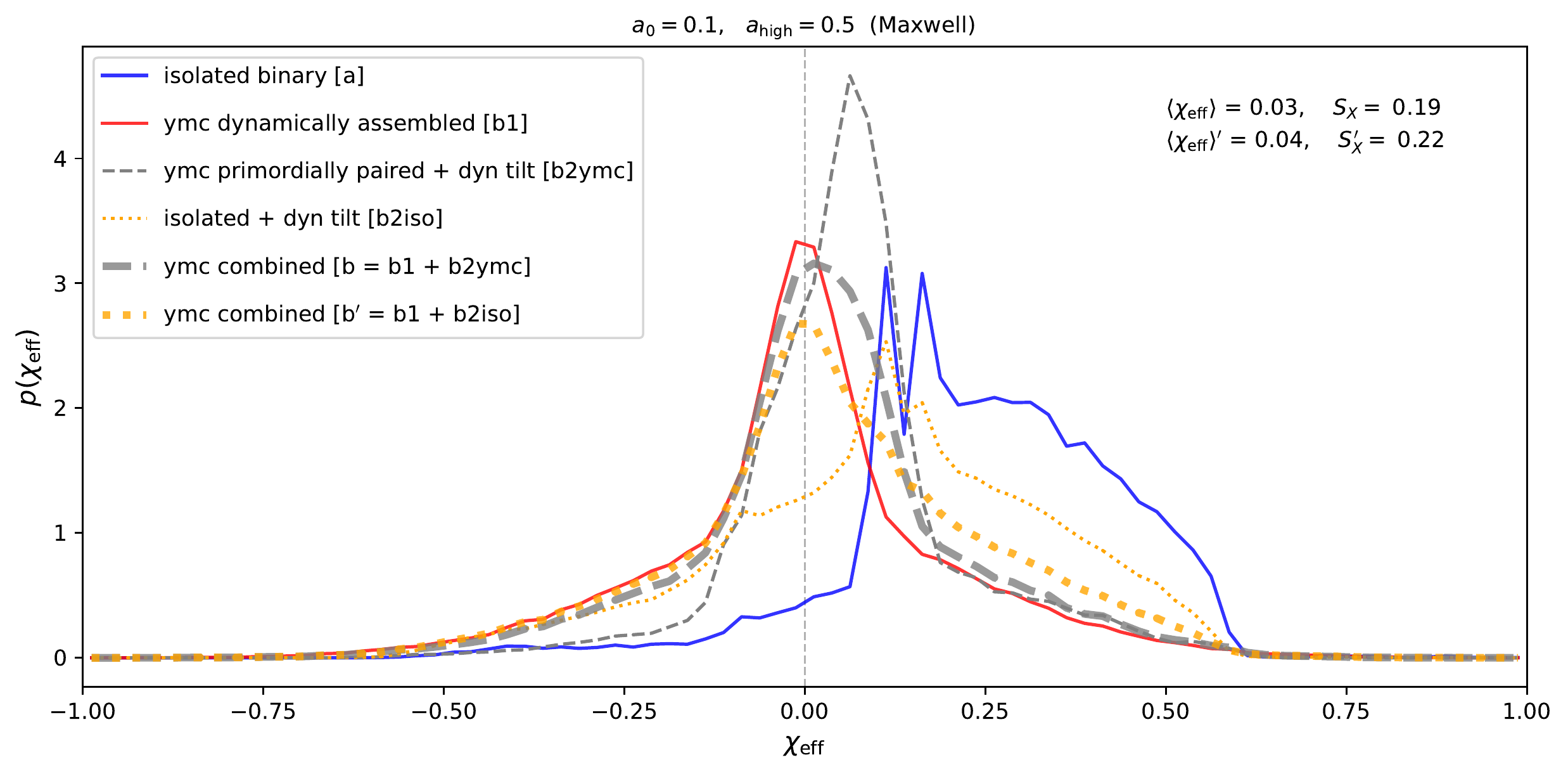}
\includegraphics[width=14.0cm]{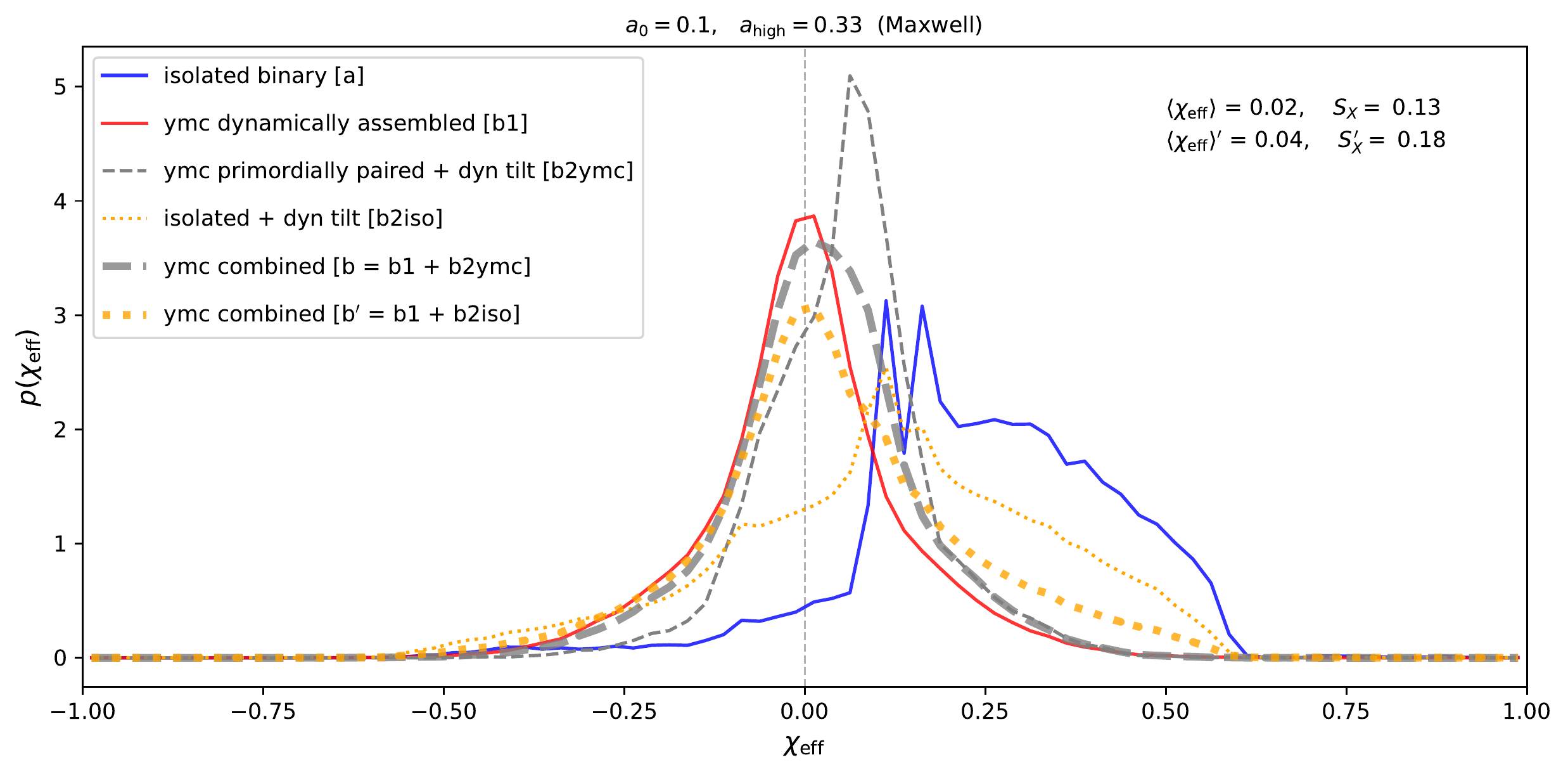}
\includegraphics[width=14.0cm]{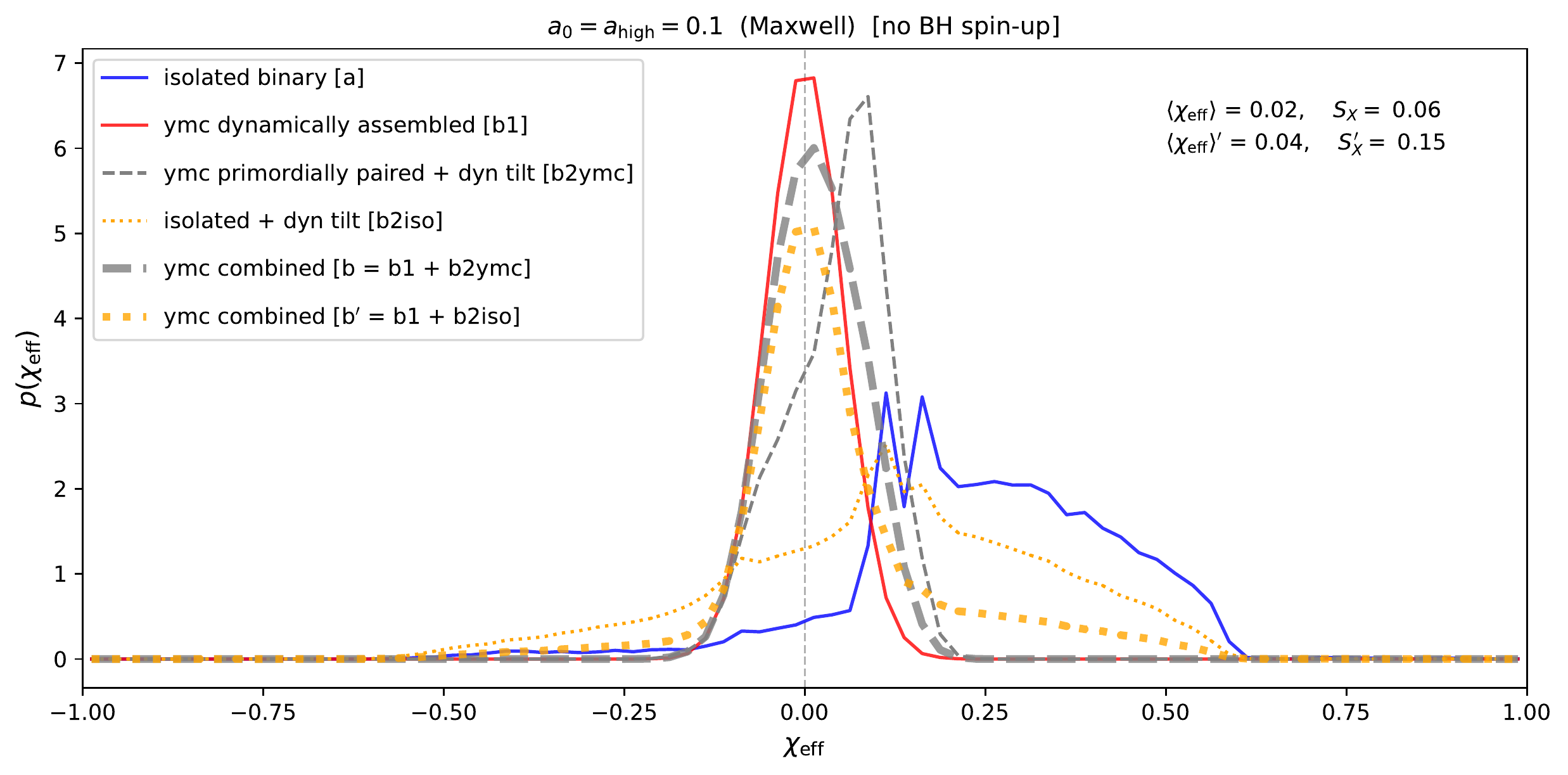}
\caption{Same description as in Fig.~\ref{fig:xeffdist1a} applies, except that the model
	isolated-binary population evolution forms BBH mergers primarily via
 	stable RLOF (the RLOF21 model).}
\label{fig:xeffdist_noce}
\end{figure*}

Figs.~\ref{fig:theta_dist_noce} and \ref{fig:xeffdist_noce}
and Table~\ref{tab:distprops_noce} show the results
corresponding to the RLOF21 isolated binary evolution model (Sec.~\ref{binmodel}).
The tilt-angle and $\xeff$ distributions and their mean values are close to
those for the CE21 isolated binary evolution case (Sec.~\ref{binmodel}; Figs.~\ref{fig:theta_dist}
and \ref{fig:xeffdist1a}).

\begin{deluxetable*}{lrrr}
\tablecaption{Statistics of the model $\xeff$ distributions for the RLOF21 isolated binary model.}
\tablewidth{0pt}
\tablehead{
\colhead{Population} & \colhead{$\xeffmean$} & \colhead{$\xeffsd$} & \colhead{$\fxneg$}
}
\startdata
	LVK observed & $0.06^{+0.03}_{-0.04}$ & $0.11^{+0.04}_{-0.03}$ & $28.32^{+14.46}_{-13.21}$ \\
\hline
	isolated [{\casea}] & 0.24 & 0.18 & 7.52 \\
	isolated + dyn. tilt [{\btwoiso}] & 0.11 & 0.22 & 28.39 \\
\hline
	\multicolumn{4}{c}{$\alow=0.1,{\rm ~~~~}\ahigh=0.5${\rm~~~~}(Maxwell)}\\
\hline
	ymc combined [{\caseb} = {\bone} + {\btwoymc}] & 0.03  & 0.19 & 41.34  \\
	ymc combined [{\casebb} = {\bone} + {\btwoiso}] & 0.04  & 0.22 & 41.73  \\
%	isolated [{\casea}] & 0.24 & 0.18 & 7.52 \\
%	isolated + dyn. tilt [{\btwoiso}] & 0.11 & 0.22 & 28.20 \\
	ymc dyn. assembled [{\bone}] & 0.00 & 0.20 & 49.85 \\ 
	ymc prim. paired + dyn. tilt [{\btwoymc}] & 0.07 & 0.15 & 27.60 \\
\hline
\multicolumn{4}{c}{$\alow=0.1,{\rm ~~~~}\ahigh=0.33${\rm~~~~}(Maxwell)}\\
\hline
	ymc combined [{\caseb} = {\bone} + {\btwoymc}] & 0.02  & 0.13 & 41.40  \\
	ymc combined [{\casebb} = {\bone} + {\btwoiso}] & 0.04  & 0.18 & 41.75  \\
%	isolated [{\casea}] & 0.24 & 0.18 & 7.52 \\
%	isolated + dyn. tilt [{\btwoiso}] & 0.11 & 0.22 & 28.39 \\
	ymc dyn. assembled [{\bone}] & 0.00 & 0.14 & 49.82 \\ 
	ymc prim. paired + dyn. tilt [{\btwoymc}] & 0.06 & 0.11 & 27.64 \\
\hline
	\multicolumn{4}{c}{$\alow=\ahigh=0.1${\rm~~~~}(Maxwell){\rm~~~~}[no BH spin-up]}\\
\hline
	ymc combined [{\caseb} = {\bone} + {\btwoymc}] & 0.02  & 0.06 & 41.31  \\
	ymc combined [{\casebb} = {\bone} + {\btwoiso}] & 0.04 & 0.15 & 41.88  \\
%	isolated [{\casea}] & 0.24 & 0.18 & 7.52 \\
%	isolated + dyn. tilt [{\btwoiso}] & 0.11 & 0.22 & 28.29 \\
	ymc dyn. assembled [{\bone}] & 0.00 & 0.05 & 50.05 \\ 
	ymc prim. paired + dyn. tilt [{\btwoymc}] & 0.04 & 0.07 & 26.08 \\
\enddata
\tablecomments{The same description as in Table~\ref{tab:distprops} applies.}
\label{tab:distprops_noce}
\end{deluxetable*}

\section{Results with additional BH-spin models}\label{altspin}

\begin{figure*}
\centering
\includegraphics[width=14.0cm]{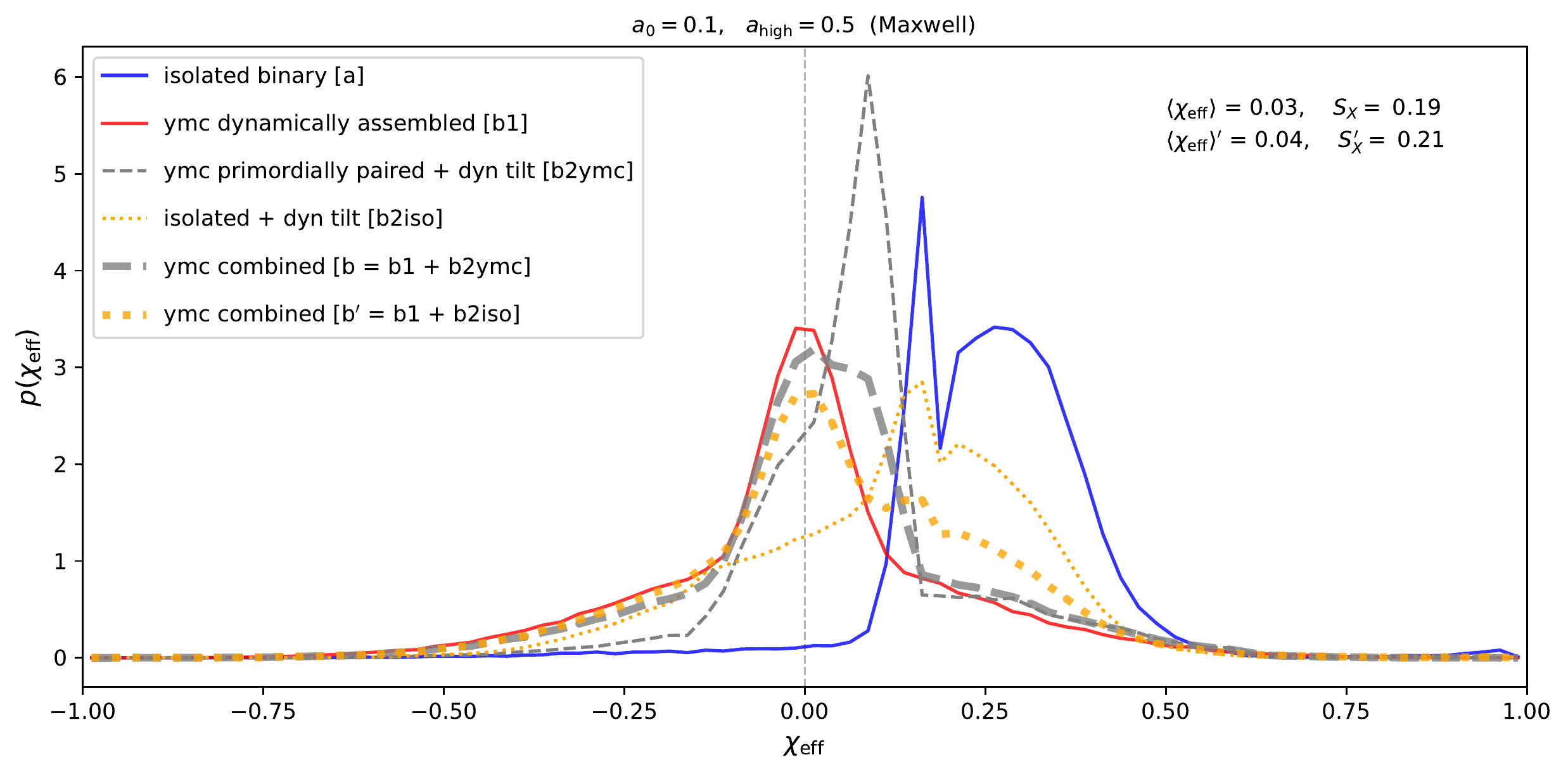}
\includegraphics[width=14.0cm]{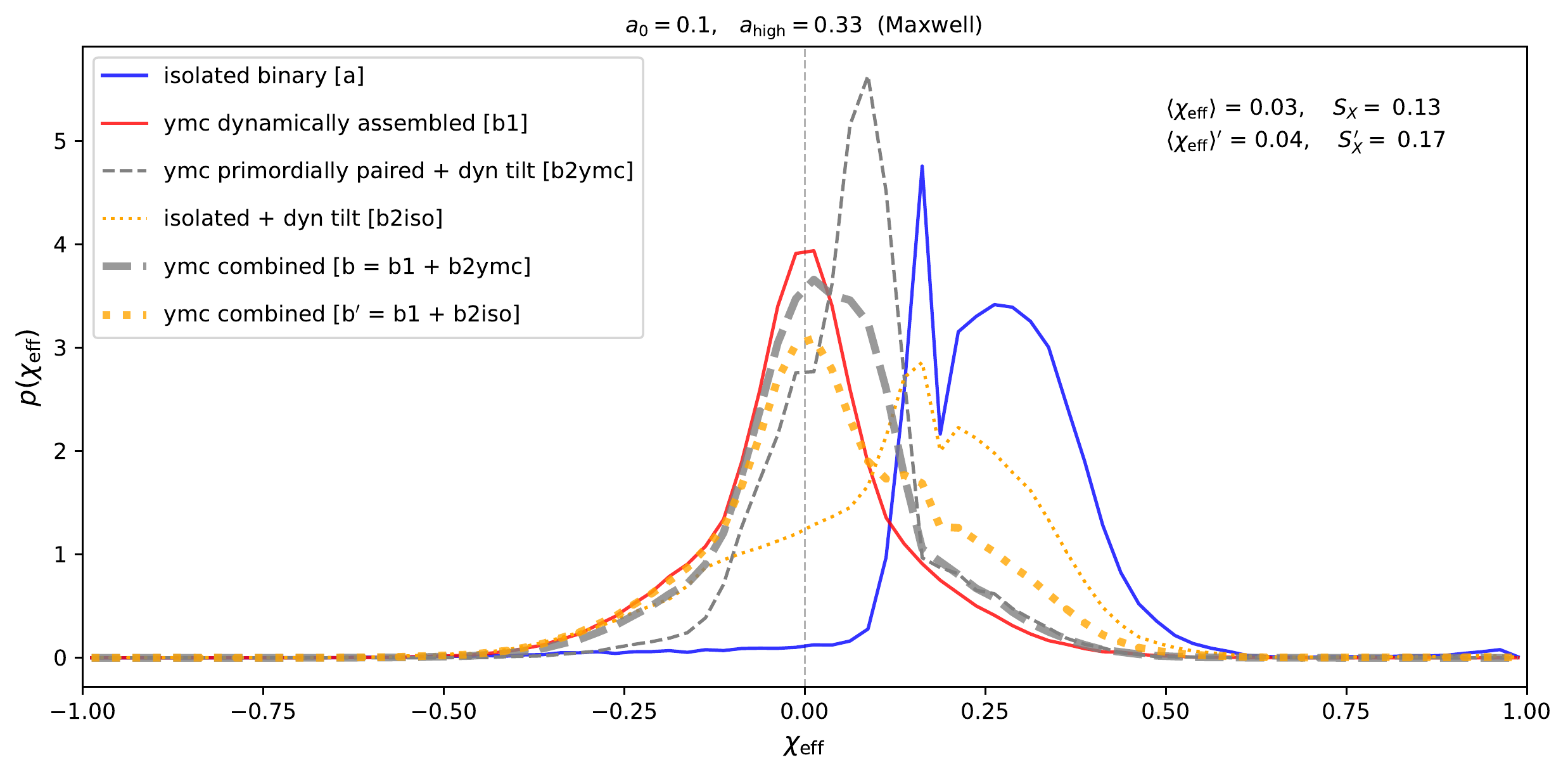}
\includegraphics[width=14.0cm]{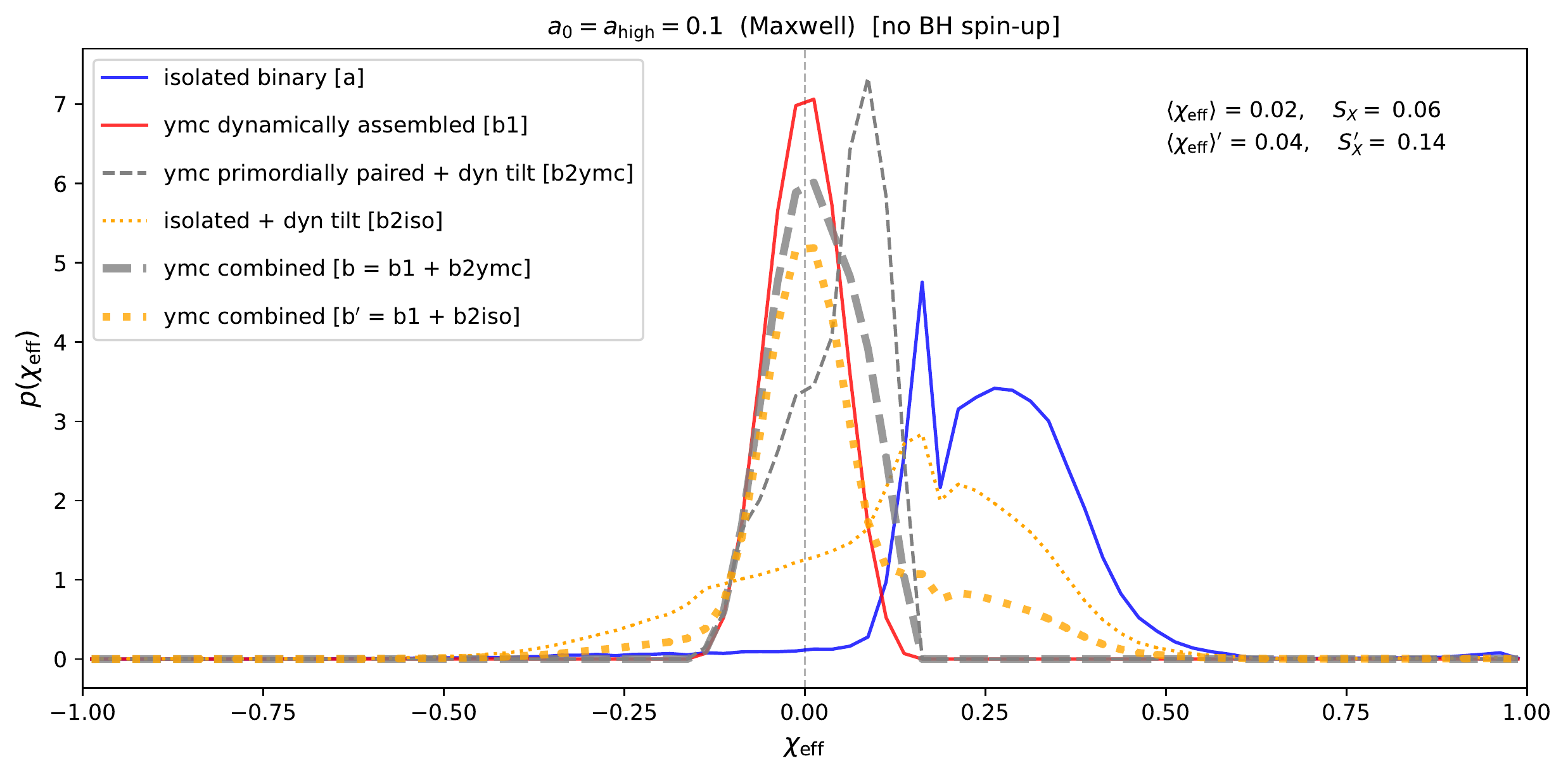}
	\caption{Same description as in Fig.~\ref{fig:xeffdist1a} applies
	(the CE21 isolated-binary model), except
	that the base BH spin is truncated to be within $\alow\in[0.05,0.15]$. This
	truncation makes the range of $\alow$ consistent with that of
	\citet[][their MESA model]{Belczynski_2020}.}
\label{fig:xeffdist_tr}
\end{figure*}

\begin{figure*}
\centering
\includegraphics[width=14.0cm]{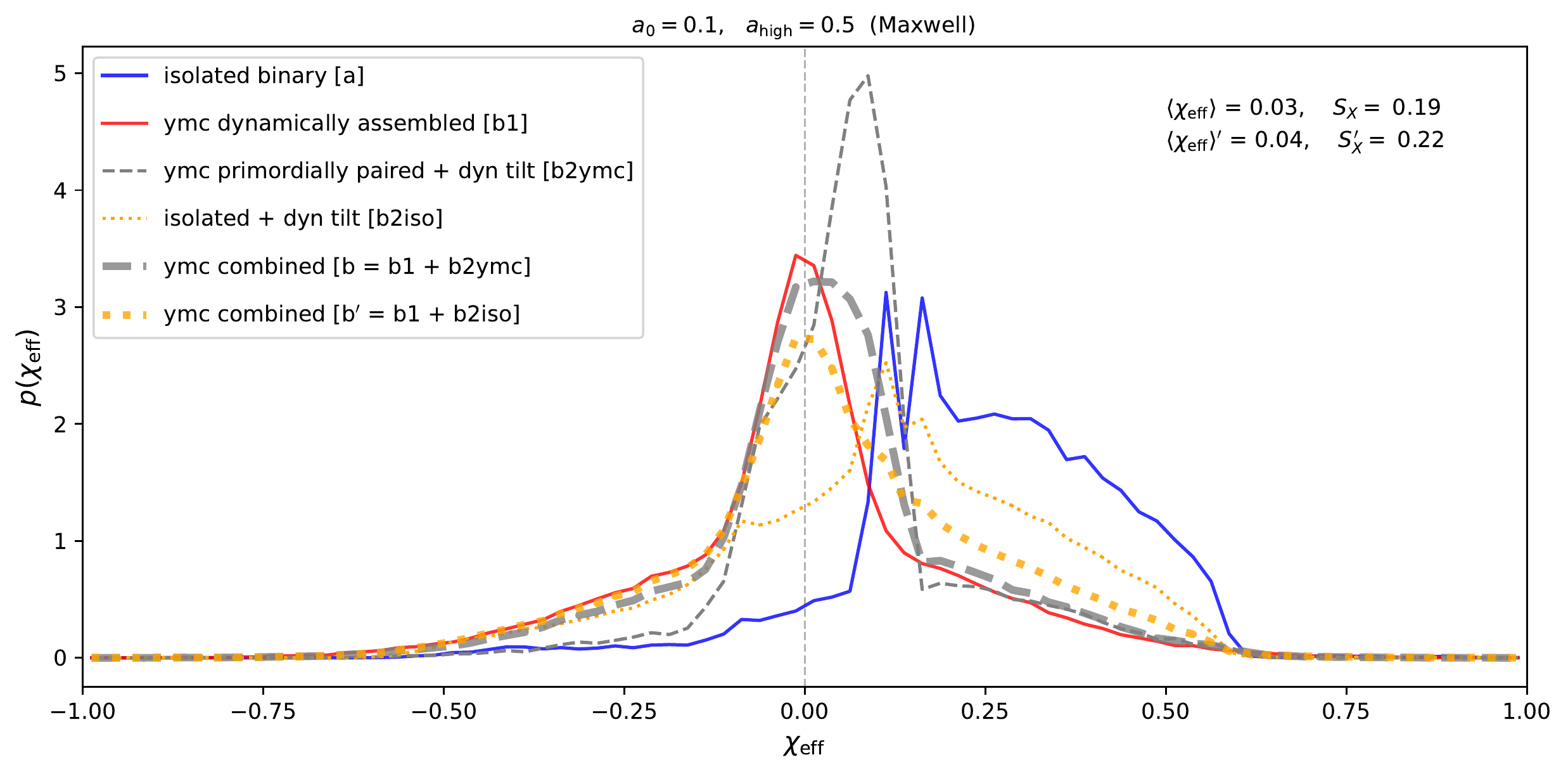}
\includegraphics[width=14.0cm]{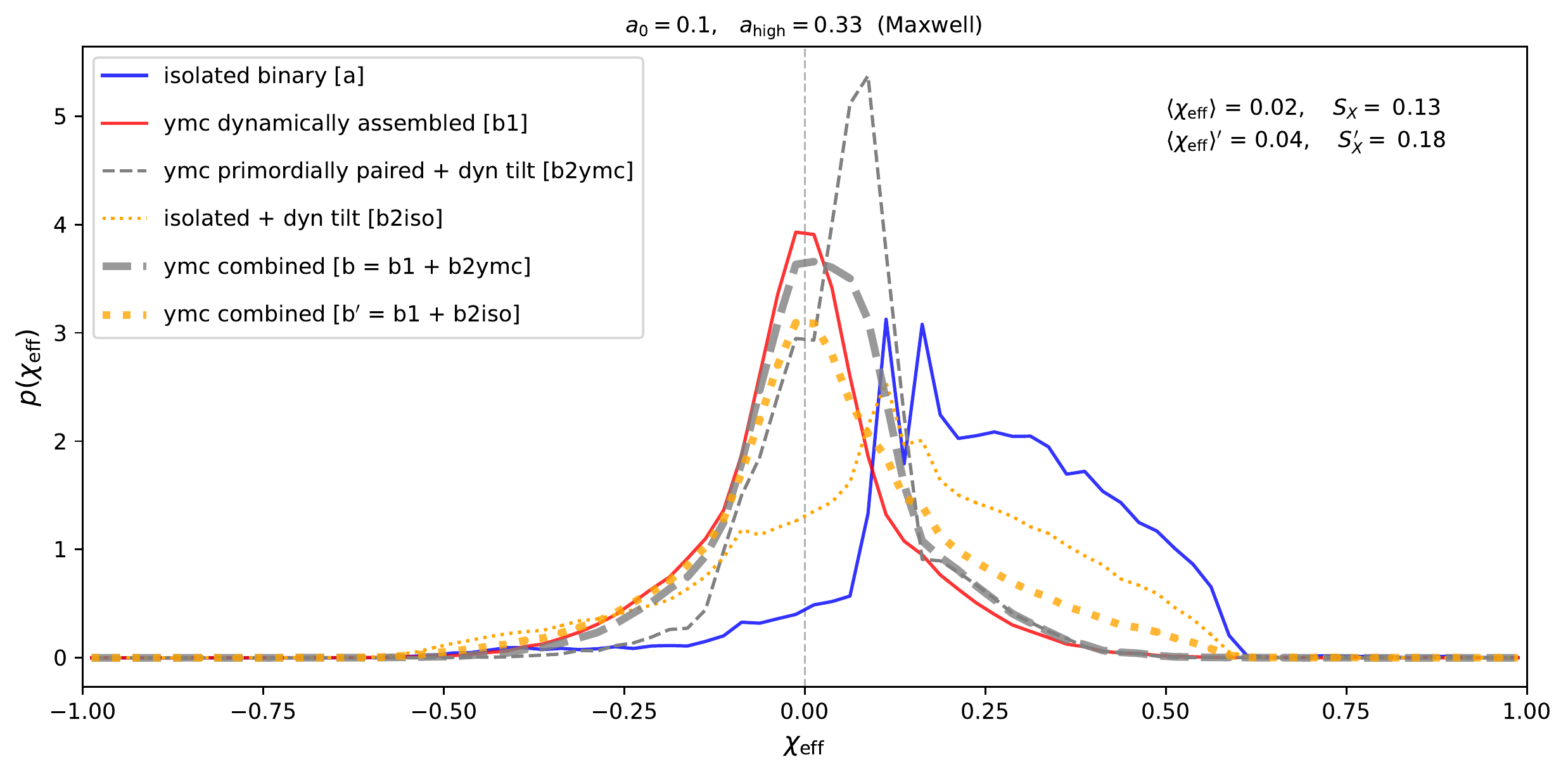}
\includegraphics[width=14.0cm]{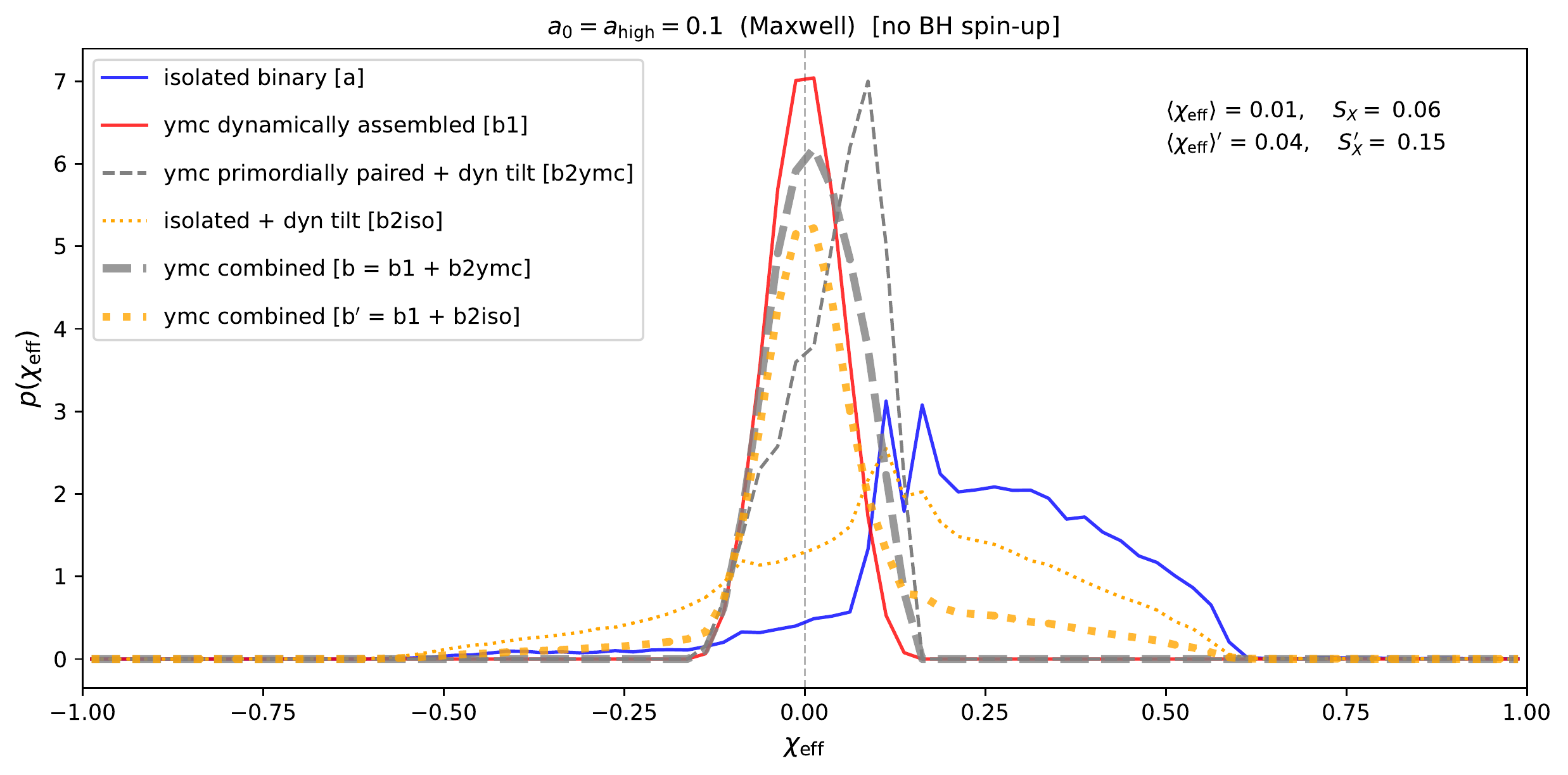}
	\caption{Same description as in Fig.~\ref{fig:xeffdist_tr} applies, except that the model
	isolated-binary population evolution forms BBH mergers primarily via
 	stable RLOF (the RLOF21 model).}
\label{fig:xeffdist_noce_tr}
\end{figure*}

\begin{figure*}
\centering
\includegraphics[width=14.0cm]{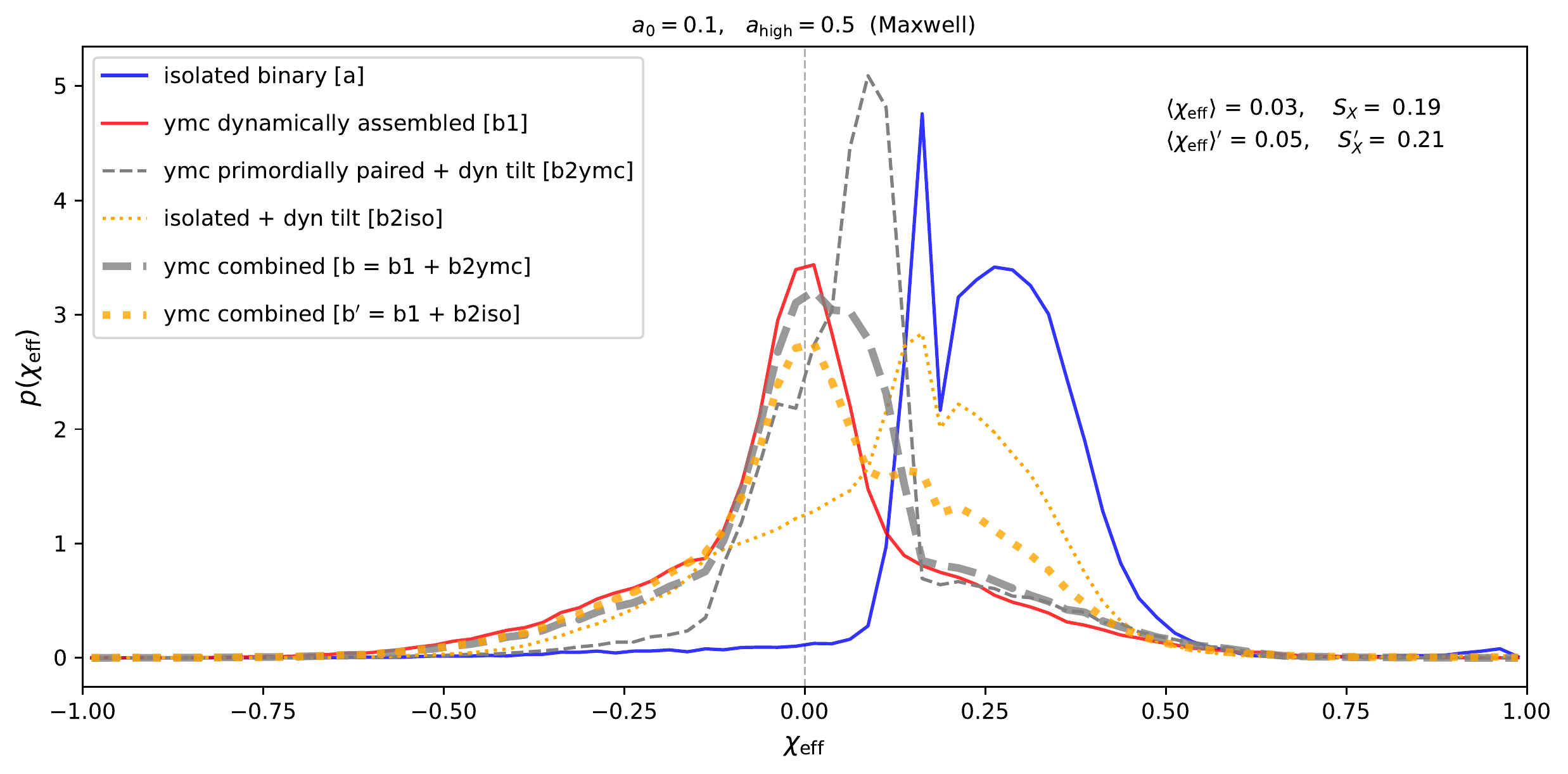}
\includegraphics[width=14.0cm]{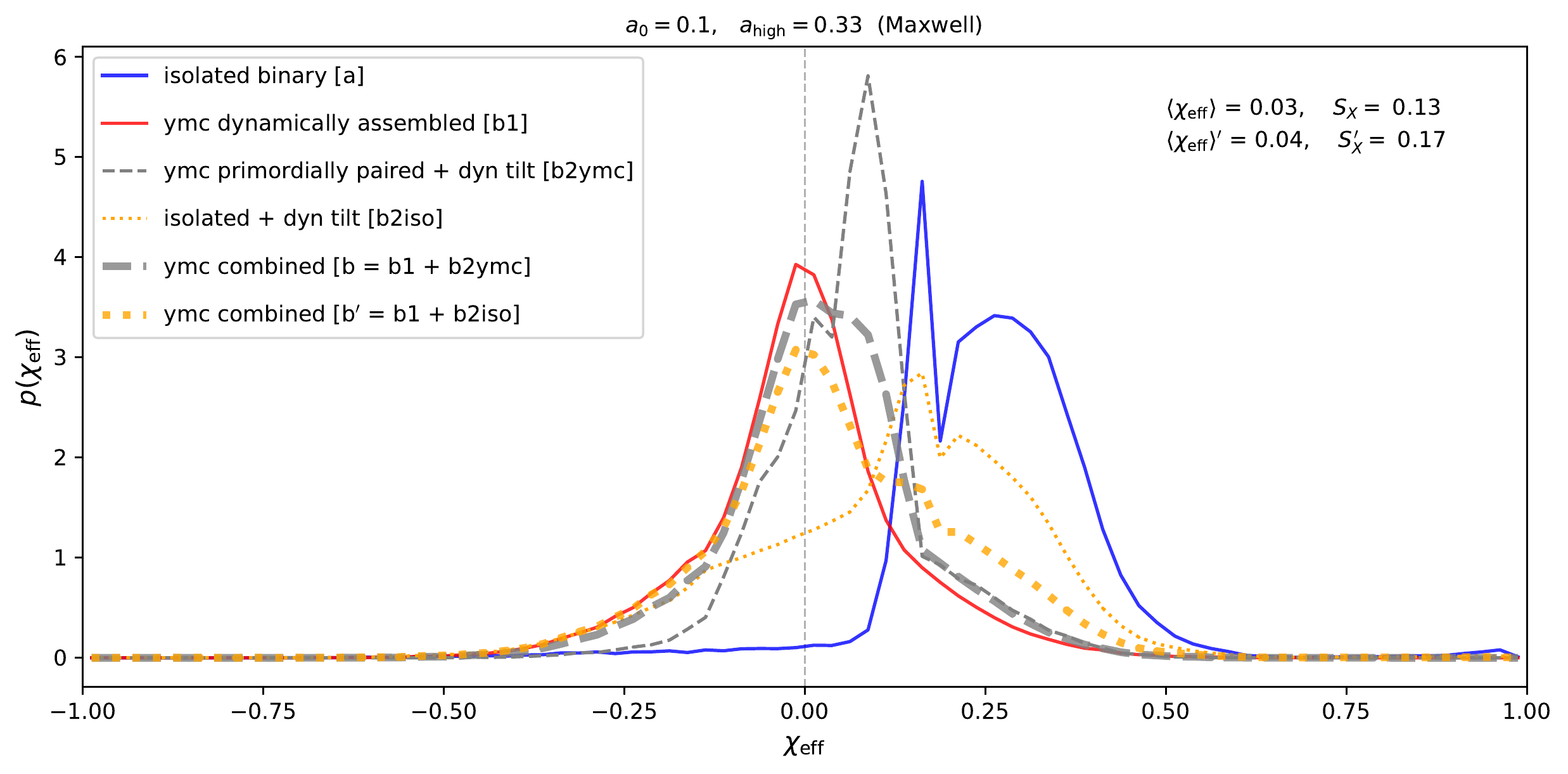}
\includegraphics[width=14.0cm]{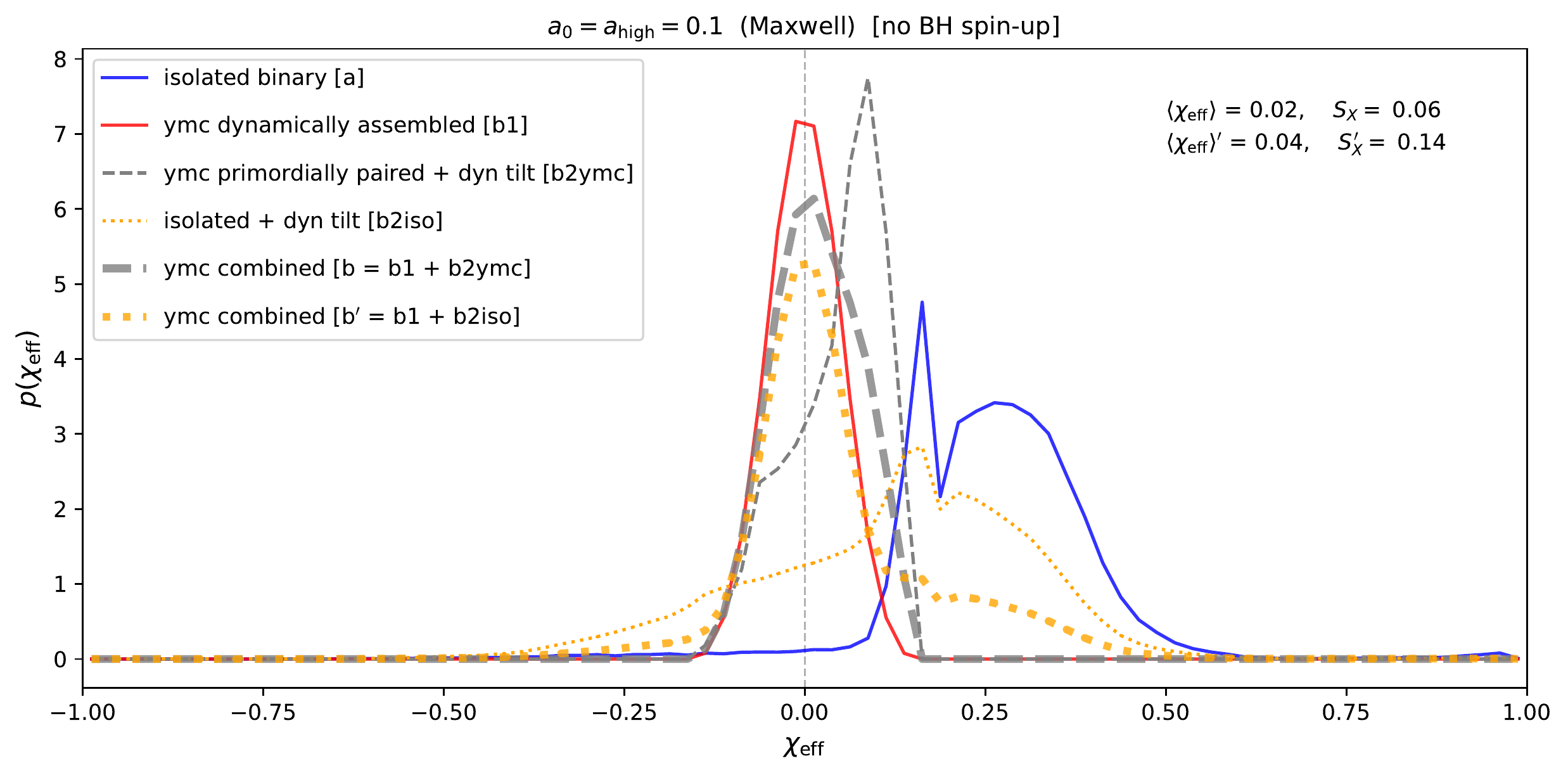}
	\caption{Same description as in Fig.~\ref{fig:xeffdist_tr} applies
	(the CE21 isolated-binary model), except
	that the spins of the spun-up BHs are truncated to be within $\ahigh\in[0.15,1.0]$,
	in addition to $\alow\in[0.05,0.15]$. These
	truncations make the ranges of $\alow$ and $\ahigh$ consistent with those of
	\citet[][]{Belczynski_2020}.}
\label{fig:xeffdist_tr3}
\end{figure*}

In this section, we show additional cases based on the choice
of BH spin. The cases are correspondingly described in the caption
of Figs.~\ref{fig:xeffdist_tr}, \ref{fig:xeffdist_noce_tr} and \ref{fig:xeffdist_tr3}.
Overall, the distribution statistics for these cases are similar to those
in Tables~\ref{tab:distprops} and \ref{tab:distprops_noce}.
In particular, $\fxneg\approx40$\% in all cases.
For reference, the statistics corresponding to the distributions in
Fig.~\ref{fig:xeffdist_tr3} are provided in Table~\ref{tab:distprops_tr3}.

%\newpage

\begin{deluxetable*}{lrrr}
\tablecaption{Statistics of the model $\xeff$ distributions corresponding to
Fig.~\ref{fig:xeffdist_tr3} (truncated $\alow$ and $\ahigh$; CE21 isolated-binary model).}
\tablewidth{0pt}
\tablehead{
\colhead{Population} & \colhead{$\xeffmean$} & \colhead{$\xeffsd$} & \colhead{$\fxneg$}
}
\startdata
	LVK observed & $0.06^{+0.03}_{-0.04}$ & $0.11^{+0.04}_{-0.03}$ & $28.32^{+14.46}_{-13.21}$ \\
\hline
	isolated [{\casea}] & 0.26 & 0.14 & 2.83 \\
	isolated + dyn. tilt [{\btwoiso}] & 0.12 & 0.19 & 25.28 \\
\hline
	\multicolumn{4}{c}{$\alow=0.1,{\rm ~~~~}\ahigh=0.5${\rm~~~~}(Maxwell)}\\
\hline
	ymc combined [{\caseb} = {\bone} + {\btwoymc}] & 0.03  & 0.19 & 40.50  \\
	ymc combined [{\casebb} = {\bone} + {\btwoiso}] & 0.05  & 0.21 & 40.65  \\
	ymc dyn. assembled [{\bone}] & 0.00 & 0.20 & 50.04 \\ 
	ymc prim. paired + dyn. tilt [{\btwoymc}] & 0.08 & 0.15 & 24.97 \\
\hline
\multicolumn{4}{c}{$\alow=0.1,{\rm ~~~~}\ahigh=0.33${\rm~~~~}(Maxwell)}\\
\hline
	ymc combined [{\caseb} = {\bone} + {\btwoymc}] & 0.03  & 0.13 & 40.66  \\
	ymc combined [{\casebb} = {\bone} + {\btwoiso}] & 0.04  & 0.17 & 40.82  \\
	ymc dyn. assembled [{\bone}] & 0.00 & 0.14 & 50.23 \\ 
	ymc prim. paired + dyn. tilt [{\btwoymc}] & 0.06 & 0.11 & 24.43 \\
\hline
	\multicolumn{4}{c}{$\alow=\ahigh=0.1${\rm~~~~}(Maxwell){\rm~~~~}[no BH spin-up]}\\
\hline
	ymc combined [{\caseb} = {\bone} + {\btwoymc}] & 0.02  & 0.06 & 40.46  \\
	ymc combined [{\casebb} = {\bone} + {\btwoiso}] & 0.04 & 0.14 & 40.69  \\
	ymc dyn. assembled [{\bone}] & 0.00 & 0.05 & 50.09 \\ 
	ymc prim. paired + dyn. tilt [{\btwoymc}] & 0.05 & 0.06 & 24.55 \\
\enddata
\tablecomments{The same description as in Table~\ref{tab:distprops} applies.}
\label{tab:distprops_tr3}
\end{deluxetable*}

\bibliography{bibliography/biblio.bib}{}

\begin{thebibliography}{}
\expandafter\ifx\csname natexlab\endcsname\relax\def\natexlab#1{#1}\fi
\providecommand{\url}[1]{\href{#1}{#1}}
\providecommand{\dodoi}[1]{doi:~\href{http://doi.org/#1}{\nolinkurl{#1}}}
\providecommand{\doeprint}[1]{\href{http://ascl.net/#1}{\nolinkurl{http://ascl.net/#1}}}
\providecommand{\doarXiv}[1]{\href{https://arxiv.org/abs/#1}{\nolinkurl{https://arxiv.org/abs/#1}}}

\bibitem[{{Aarseth}(2012)}]{Aarseth_2012}
{Aarseth}, S.~J. 2012, \mnras, 422, 841,
  \dodoi{10.1111/j.1365-2966.2012.20666.x}

\bibitem[{Aasi {et~al.}(2015)Aasi, Abbott, Abbott, Abbott, Abernathy, Ackley,
  Adams, Adams, \& et~al.}]{Asai_2015}
Aasi, J., Abbott, B.~P., Abbott, R., {et~al.} 2015, Classical and Quantum
  Gravity, 32, 074001, \dodoi{10.1088/0264-9381/32/7/074001}

\bibitem[{{Abbott} {et~al.}(2019){Abbott}, {Abbott}, {Abbott}, {Abraham},
  {Acernese}, {Ackley}, {Adams}, {Adhikari}, {Adya}, {Affeldt}, \&
  et~al.}]{Abbott_GWTC1}
{Abbott}, B.~P., {Abbott}, R., {Abbott}, T.~D., {et~al.} 2019, Physical Review
  X, 9, 031040, \dodoi{10.1103/PhysRevX.9.031040}

\bibitem[{{Abbott} {et~al.}(2021{\natexlab{a}}){Abbott}, {Abbott}, {Abraham},
  {Acernese}, {Ackley}, {Adams}, {Adams}, {Adhikari}, \& et~al.}]{Abbott_GWTC2}
{Abbott}, R., {Abbott}, T.~D., {Abraham}, S., {et~al.} 2021{\natexlab{a}},
  Phys. Rev. X, 11, 021053, \dodoi{10.1103/PhysRevX.11.021053}

\bibitem[{{Abbott} {et~al.}(2021{\natexlab{b}}){Abbott}, {Abbott}, {Acernese},
  {Ackley}, {Adams}, {Adhikari}, {Adhikari}, {Adya}, \& et~al.}]{Abbott_GWTC3}
{Abbott}, R., {Abbott}, T.~D., {Acernese}, F., {et~al.} 2021{\natexlab{b}},
  arXiv e-prints, arXiv:2111.03606.
\newblock \doarXiv{2111.03606}

\bibitem[{{Abbott} {et~al.}(2021{\natexlab{c}}){Abbott}, {Abbott}, {Acernese},
  {Ackley}, {Adams}, {Adhikari}, {Adhikari}, {Adya}, \&
  et~al.}]{Abbott_GWTC2.1}
---. 2021{\natexlab{c}}, arXiv e-prints, arXiv:2108.01045.
\newblock \doarXiv{2108.01045}

\bibitem[{{Abbott} {et~al.}(2023){Abbott}, {Abbott}, {Acernese}, {Ackley},
  {Adams}, {Adhikari}, {Adhikari}, {Adya}, \& et~al.}]{Abbott_GWTC3_prop}
---. 2023, Physical Review X, 13, 011048, \dodoi{10.1103/PhysRevX.13.011048}

\bibitem[{{Acernese} {et~al.}(2015){Acernese}, {Agathos}, {Agatsuma}, {Aisa},
  {Allemandou}, {Allocca}, {Amarni}, {Astone}, {Balestri}, {Ballardin}, \&
  et~al.}]{Acernese_2015}
{Acernese}, F., {Agathos}, M., {Agatsuma}, K., {et~al.} 2015, Classical and
  Quantum Gravity, 32, 024001, \dodoi{10.1088/0264-9381/32/2/024001}

\bibitem[{Ajith {et~al.}(2011)Ajith, Hannam, Husa, Chen, Br\"ugmann, Dorband,
  M\"uller, Ohme, Pollney, Reisswig, Santamar\'{\i}a, \& Seiler}]{Ajith_2011}
Ajith, P., Hannam, M., Husa, S., {et~al.} 2011, Phys. Rev. Lett., 106, 241101,
  \dodoi{10.1103/PhysRevLett.106.241101}

\bibitem[{{Antonini} \& {Gieles}(2020)}]{Antonini_2020b}
{Antonini}, F., \& {Gieles}, M. 2020, \prd, 102, 123016,
  \dodoi{10.1103/PhysRevD.102.123016}

\bibitem[{{Arca Sedda} {et~al.}(2021){Arca Sedda}, {Mapelli}, {Benacquista}, \&
  {Spera}}]{ArcaSedda_2021b}
{Arca Sedda}, M., {Mapelli}, M., {Benacquista}, M., \& {Spera}, M. 2021, arXiv
  e-prints, arXiv:2109.12119.
\newblock \doarXiv{2109.12119}

\bibitem[{{Arca Sedda} {et~al.}(2020){Arca Sedda}, {Mapelli}, {Spera},
  {Benacquista}, \& {Giacobbo}}]{ArcaSedda_2020b}
{Arca Sedda}, M., {Mapelli}, M., {Spera}, M., {Benacquista}, M., \& {Giacobbo},
  N. 2020, \apj, 894, 133, \dodoi{10.3847/1538-4357/ab88b2}

\bibitem[{Banerjee(2020)}]{Banerjee_2020d}
Banerjee, S. 2020, Phys. Rev. D, 102, 103002,
  \dodoi{10.1103/PhysRevD.102.103002}

\bibitem[{{Banerjee}(2021{\natexlab{a}})}]{Banerjee_2020c}
{Banerjee}, S. 2021{\natexlab{a}}, \mnras, 500, 3002,
  \dodoi{10.1093/mnras/staa2392}

\bibitem[{{Banerjee}(2021{\natexlab{b}})}]{Banerjee_2021}
---. 2021{\natexlab{b}}, \mnras, 503, 3371, \dodoi{10.1093/mnras/stab591}

\bibitem[{{Banerjee}(2022{\natexlab{a}})}]{Banerjee_2022}
---. 2022{\natexlab{a}}, \aap, 665, A20, \dodoi{10.1051/0004-6361/202142331}

\bibitem[{{Banerjee}(2022{\natexlab{b}})}]{Banerjee_2021b}
---. 2022{\natexlab{b}}, \prd, 105, 023004, \dodoi{10.1103/PhysRevD.105.023004}

\bibitem[{{Banerjee} {et~al.}(2020){Banerjee}, {Belczynski}, {Fryer},
  {Berczik}, {Hurley}, {Spurzem}, \& {Wang}}]{Banerjee_2020}
{Banerjee}, S., {Belczynski}, K., {Fryer}, C.~L., {et~al.} 2020, \aap, 639,
  A41, \dodoi{10.1051/0004-6361/201935332}

\bibitem[{{Bavera} {et~al.}(2022){Bavera}, {Fishbach}, {Zevin}, {Zapartas}, \&
  {Fragos}}]{Bavera_2022}
{Bavera}, S.~S., {Fishbach}, M., {Zevin}, M., {Zapartas}, E., \& {Fragos}, T.
  2022, A\&A, 665, A59, \dodoi{10.1051/0004-6361/202243724}

\bibitem[{{Belczynski} {et~al.}(2022{\natexlab{a}}){Belczynski}, {Doctor},
  {Zevin}, {Olejak}, {Banerje}, \& {Chattopadhyay}}]{Belczynski_2022}
{Belczynski}, K., {Doctor}, Z., {Zevin}, M., {et~al.} 2022{\natexlab{a}}, \apj,
  935, 126, \dodoi{10.3847/1538-4357/ac8167}

\bibitem[{Belczynski {et~al.}(2008)Belczynski, Kalogera, Rasio, Taam, Zezas,
  Bulik, Maccarone, \& Ivanova}]{Belczynski_2008}
Belczynski, K., Kalogera, V., Rasio, F.~A., {et~al.} 2008, The Astrophysical
  Journal Supplement Series, 174, 223, \dodoi{10.1086/521026}

\bibitem[{{Belczynski} {et~al.}(2010){Belczynski}, {Lorimer}, {Ridley}, \&
  {Curran}}]{Belczynski_2010b}
{Belczynski}, K., {Lorimer}, D.~R., {Ridley}, J.~P., \& {Curran}, S.~J. 2010,
  \mnras, 407, 1245, \dodoi{10.1111/j.1365-2966.2010.16970.x}

\bibitem[{{Belczynski} {et~al.}(2016){Belczynski}, {Heger}, {Gladysz},
  {Ruiter}, {Woosley}, {Wiktorowicz}, {Chen}, {Bulik}, {O'Shaughnessy}, {Holz},
  {Fryer}, \& {Berti}}]{Belczynski_2016a}
{Belczynski}, K., {Heger}, A., {Gladysz}, W., {et~al.} 2016, \aap, 594, A97,
  \dodoi{10.1051/0004-6361/201628980}

\bibitem[{{Belczynski} {et~al.}(2020){Belczynski}, {Klencki}, {Fields},
  {Olejak}, {Berti}, {Meynet}, {Fryer}, {Holz}, {O'Shaughnessy}, {Brown},
  {Bulik}, {Leung}, {Nomoto}, {Madau}, {Hirschi}, {Kaiser}, {Jones}, {Mondal},
  {Chruslinska}, {Drozda}, {Gerosa}, {Doctor}, {Giersz}, {Ekstrom}, {Georgy},
  {Askar}, {Baibhav}, {Wysocki}, {Natan}, {Farr}, {Wiktorowicz}, {Coleman
  Miller}, {Farr}, \& {Lasota}}]{Belczynski_2020}
{Belczynski}, K., {Klencki}, J., {Fields}, C.~E., {et~al.} 2020, \aap, 636,
  A104, \dodoi{10.1051/0004-6361/201936528}

\bibitem[{{Belczynski} {et~al.}(2022{\natexlab{b}}){Belczynski}, {Romagnolo},
  {Olejak}, {Klencki}, {Chattopadhyay}, {Stevenson}, {Coleman Miller},
  {Lasota}, \& {Crowther}}]{Belczynski_2022a}
{Belczynski}, K., {Romagnolo}, A., {Olejak}, A., {et~al.} 2022{\natexlab{b}},
  \apj, 925, 69, \dodoi{10.3847/1538-4357/ac375a}

\bibitem[{{Branchesi}(2016)}]{Branchesi_2016}
{Branchesi}, M. 2016, in Journal of Physics Conference Series, Vol. 718,
  Journal of Physics Conference Series, 022004,
  \dodoi{10.1088/1742-6596/718/2/022004}

\bibitem[{{Briel} {et~al.}(2021){Briel}, {Eldridge}, {Stanway}, {Stevance}, \&
  {Chrimes}}]{Briel_2021}
{Briel}, M.~M., {Eldridge}, J.~J., {Stanway}, E.~R., {Stevance}, H.~F., \&
  {Chrimes}, A.~A. 2021, arXiv e-prints, arXiv:2111.08124.
\newblock \doarXiv{2111.08124}

\bibitem[{{Broekgaarden} {et~al.}(2022){Broekgaarden}, {Stevenson}, \&
  {Thrane}}]{Broekgaarden2022}
{Broekgaarden}, F.~S., {Stevenson}, S., \& {Thrane}, E. 2022, \apj, 938, 45,
  \dodoi{10.3847/1538-4357/ac8879}

\bibitem[{{Callister} \& {Farr}(2023)}]{Callister2023}
{Callister}, T.~A., \& {Farr}, W.~M. 2023, arXiv e-prints, arXiv:2302.07289,
  \dodoi{10.48550/arXiv.2302.07289}

\bibitem[{{Callister} {et~al.}(2022){Callister}, {Miller}, {Chatziioannou}, \&
  {Farr}}]{Callister2022}
{Callister}, T.~A., {Miller}, S.~J., {Chatziioannou}, K., \& {Farr}, W.~M.
  2022, \apjl, 937, L13, \dodoi{10.3847/2041-8213/ac847e}

\bibitem[{{Carole} {et~al.}(2023){Carole}, {Michela}, {Filippo}, {Yann}, \&
  {Roberta}}]{Perigois_2023}
{Carole}, P., {Michela}, M., {Filippo}, S., {Yann}, B., \& {Roberta}, R. 2023,
  arXiv e-prints, arXiv:2301.01312.
\newblock \doarXiv{2301.01312}

\bibitem[{{Chatterjee} {et~al.}(2017){Chatterjee}, {Rodriguez}, \&
  {Rasio}}]{Chatterjee_2017a}
{Chatterjee}, S., {Rodriguez}, C.~L., \& {Rasio}, F.~A. 2017, \apj, 834, 68,
  \dodoi{10.3847/1538-4357/834/1/68}

\bibitem[{{Chattopadhyay} {et~al.}(2022){Chattopadhyay}, {Hurley}, {Stevenson},
  \& {Raidani}}]{Chatto_2022}
{Chattopadhyay}, D., {Hurley}, J., {Stevenson}, S., \& {Raidani}, A. 2022,
  \mnras, 513, 4527, \dodoi{10.1093/mnras/stac1163}

\bibitem[{{Chru{\'s}li{\'n}ska}(2022)}]{Chruslinska_2022}
{Chru{\'s}li{\'n}ska}, M. 2022, arXiv e-prints, arXiv:2206.10622.
\newblock \doarXiv{2206.10622}

\bibitem[{{Chruslinska} \& {Nelemans}(2019)}]{Chruslinska_2019}
{Chruslinska}, M., \& {Nelemans}, G. 2019, \mnras, 488, 5300,
  \dodoi{10.1093/mnras/stz2057}

\bibitem[{{de Mink} {et~al.}(2010){de Mink}, {Cantiello}, {Langer}, \&
  {Pols}}]{deMink_2010}
{de Mink}, S.~E., {Cantiello}, M., {Langer}, N., \& {Pols}, O.~R. 2010, in
  American Institute of Physics Conference Series, Vol. 1314, International
  Conference on Binaries: in celebration of Ron Webbink's 65th Birthday, ed.
  V.~{Kalogera} \& M.~{van der Sluys}, 291--296, \dodoi{10.1063/1.3536387}

\bibitem[{{De Mink} {et~al.}(2009){De Mink}, {Cantiello}, {Langer}, {Pols},
  {Brott}, \& {Yoon}}]{deMink_2009}
{De Mink}, S.~E., {Cantiello}, M., {Langer}, N., {et~al.} 2009, \aap, 497, 243,
  \dodoi{10.1051/0004-6361/200811439}

\bibitem[{{De Mink} \& {Mandel}(2016)}]{DeMink_2016}
{De Mink}, S.~E., \& {Mandel}, I. 2016, \mnras, 460, 3545,
  \dodoi{10.1093/mnras/stw1219}

\bibitem[{{Di Carlo} {et~al.}(2020){Di Carlo}, {Mapelli}, {Giacobbo}, {Spera},
  {Bouffanais}, {Rastello}, {Santoliquido}, {Pasquato}, {Ballone}, {Trani},
  {Torniamenti}, \& {Haardt}}]{DiCarlo_2020}
{Di Carlo}, U.~N., {Mapelli}, M., {Giacobbo}, N., {et~al.} 2020, \mnras, 498,
  495, \dodoi{10.1093/mnras/staa2286}

\bibitem[{{Fryer} {et~al.}(2012){Fryer}, {Belczynski}, {Wiktorowicz},
  {Dominik}, {Kalogera}, \& {Holz}}]{Fryer_2012}
{Fryer}, C.~L., {Belczynski}, K., {Wiktorowicz}, G., {et~al.} 2012, \apj, 749,
  91, \dodoi{10.1088/0004-637X/749/1/91}

\bibitem[{{Fryer} {et~al.}(2022){Fryer}, {Olejak}, \&
  {Belczynski}}]{Fryer_2022}
{Fryer}, C.~L., {Olejak}, A., \& {Belczynski}, K. 2022, \apj, 931, 94,
  \dodoi{10.3847/1538-4357/ac6ac9}

\bibitem[{{Fuller} \& {Lu}(2022)}]{Fuller_2022}
{Fuller}, J., \& {Lu}, W. 2022, \mnras, 511, 3951,
  \dodoi{10.1093/mnras/stac317}

\bibitem[{{Fuller} \& {Ma}(2019)}]{Fuller_2019a}
{Fuller}, J., \& {Ma}, L. 2019, \apjl, 881, L1,
  \dodoi{10.3847/2041-8213/ab339b}

\bibitem[{{Galaudage} {et~al.}(2021){Galaudage}, {Talbot}, {Nagar}, {Jain},
  {Thrane}, \& {Mandel}}]{Galaudage2021}
{Galaudage}, S., {Talbot}, C., {Nagar}, T., {et~al.} 2021, \apjl, 921, L15,
  \dodoi{10.3847/2041-8213/ac2f3c}

\bibitem[{{Geller} {et~al.}(2013){Geller}, {Hurley}, \&
  {Mathieu}}]{Geller_2013}
{Geller}, A.~M., {Hurley}, J.~R., \& {Mathieu}, R.~D. 2013, \aj, 145, 8,
  \dodoi{10.1088/0004-6256/145/1/8}

\bibitem[{{Gerosa} \& {Berti}(2017)}]{Gerosa_2017}
{Gerosa}, D., \& {Berti}, E. 2017, \prd, 95, 124046,
  \dodoi{10.1103/PhysRevD.95.124046}

\bibitem[{{Gerosa} \& {Fishbach}(2021)}]{Gerosa_2021}
{Gerosa}, D., \& {Fishbach}, M. 2021, Nature Astronomy, 5, 749,
  \dodoi{10.1038/s41550-021-01398-w}

\bibitem[{Harris {et~al.}(2020)Harris, Millman, van~der Walt, Gommers,
  Virtanen, Cournapeau, Wieser, Taylor, Berg, Smith, Kern, Picus, Hoyer, van
  Kerkwijk, Brett, Haldane, del R{\'{i}}o, Wiebe, Peterson,
  G{\'{e}}rard-Marchant, Sheppard, Reddy, Weckesser, Abbasi, Gohlke, \&
  Oliphant}]{harris2020array}
Harris, C.~R., Millman, K.~J., van~der Walt, S.~J., {et~al.} 2020, Nature, 585,
  357, \dodoi{10.1038/s41586-020-2649-2}

\bibitem[{{Hobbs} {et~al.}(2005){Hobbs}, {Lorimer}, {Lyne}, \&
  {Kramer}}]{Hobbs_2005}
{Hobbs}, G., {Lorimer}, D.~R., {Lyne}, A.~G., \& {Kramer}, M. 2005, \mnras,
  360, 974, \dodoi{10.1111/j.1365-2966.2005.09087.x}

\bibitem[{Hunter(2007)}]{Hunter_2007}
Hunter, J.~D. 2007, Computing in Science \& Engineering, 9, 90,
  \dodoi{10.1109/MCSE.2007.55}

\bibitem[{Hurley {et~al.}(2000)Hurley, Pols, \& Tout}]{Hurley_2000}
Hurley, J.~R., Pols, O.~R., \& Tout, C.~A. 2000, Monthly Notices of the Royal
  Astronomical Society, 315, 543, \dodoi{10.1046/j.1365-8711.2000.03426.x}

\bibitem[{Hurley {et~al.}(2002)Hurley, Tout, \& Pols}]{Hurley_2002}
Hurley, J.~R., Tout, C.~A., \& Pols, O.~R. 2002, Monthly Notices of the Royal
  Astronomical Society, 329, 897, \dodoi{10.1046/j.1365-8711.2002.05038.x}

\bibitem[{{KAGRA Collaboration} {et~al.}(2020){KAGRA Collaboration}, {Akutsu},
  {Ando}, {Arai}, {Arai}, {Araki}, {Araya}, {Aritomi}, {Asada}, {Aso}, {Bae},
  \& et~al.}]{KAGRA_2020}
{KAGRA Collaboration}, {Akutsu}, T., {Ando}, M., {et~al.} 2020, Progress of
  Theoretical and Experimental Physics, 2021, \dodoi{10.1093/ptep/ptaa120}

\bibitem[{{King} {et~al.}(2001){King}, {Davies}, {Ward}, {Fabbiano}, \&
  {Elvis}}]{King_2001}
{King}, A.~R., {Davies}, M.~B., {Ward}, M.~J., {Fabbiano}, G., \& {Elvis}, M.
  2001, \apjl, 552, L109, \dodoi{10.1086/320343}

\bibitem[{{King}(1966)}]{King_1966}
{King}, I.~R. 1966, \aj, 71, 64, \dodoi{10.1086/109857}

\bibitem[{{Kroupa}(2001)}]{Kroupa_2001}
{Kroupa}, P. 2001, \mnras, 322, 231, \dodoi{10.1046/j.1365-8711.2001.04022.x}

\bibitem[{{Krumholz} {et~al.}(2019){Krumholz}, {McKee}, \& {Bland
  -Hawthorn}}]{Krumholz_2019}
{Krumholz}, M.~R., {McKee}, C.~F., \& {Bland -Hawthorn}, J. 2019, \araa, 57,
  227, \dodoi{10.1146/annurev-astro-091918-104430}

\bibitem[{{Kumamoto} {et~al.}(2020){Kumamoto}, {Fujii}, \&
  {Tanikawa}}]{Kumamoto_2020}
{Kumamoto}, J., {Fujii}, M.~S., \& {Tanikawa}, A. 2020, \mnras, 495, 4268,
  \dodoi{10.1093/mnras/staa1440}

\bibitem[{{Ma} \& {Fuller}(2023)}]{Ma_2023}
{Ma}, L., \& {Fuller}, J. 2023, arXiv e-prints, arXiv:2305.08356,
  \dodoi{10.48550/arXiv.2305.08356}

\bibitem[{{MacLeod} {et~al.}(2017){MacLeod}, {Macias}, {Ramirez-Ruiz},
  {Grindlay}, {Batta}, \& {Montes}}]{MacLeod_2017}
{MacLeod}, M., {Macias}, P., {Ramirez-Ruiz}, E., {et~al.} 2017, \apj, 835, 282,
  \dodoi{10.3847/1538-4357/835/2/282}

\bibitem[{{Madau} \& {Fragos}(2017)}]{Madau_2017}
{Madau}, P., \& {Fragos}, T. 2017, \apj, 840, 39,
  \dodoi{10.3847/1538-4357/aa6af9}

\bibitem[{{Maeder}(1987)}]{Maeder_1987}
{Maeder}, A. 1987, A\&A, 178, 159

\bibitem[{{Mandel}(2016)}]{Mandel_2016}
{Mandel}, I. 2016, \mnras, 456, 578, \dodoi{10.1093/mnras/stv2733}

\bibitem[{{Mandel} \& {Broekgaarden}(2022)}]{Mandel_2021}
{Mandel}, I., \& {Broekgaarden}, F.~S. 2022, Living Reviews in Relativity, 25,
  1, \dodoi{10.1007/s41114-021-00034-3}

\bibitem[{{Mandel} {et~al.}(2021){Mandel}, {M{\"u}ller}, {Riley}, {de Mink},
  {Vigna-G{\'o}mez}, \& {Chattopadhyay}}]{Mandel_2021a}
{Mandel}, I., {M{\"u}ller}, B., {Riley}, J., {et~al.} 2021, \mnras, 500, 1380,
  \dodoi{10.1093/mnras/staa3390}

\bibitem[{{Mapelli}(2018)}]{Mapelli_2018}
{Mapelli}, M. 2018, in Journal of Physics Conference Series, Vol. 957, Journal
  of Physics Conference Series, 012001, \dodoi{10.1088/1742-6596/957/1/012001}

\bibitem[{{Marchant} {et~al.}(2016){Marchant}, {Langer}, {Podsiadlowski},
  {Tauris}, \& {Moriya}}]{Marchant_2016}
{Marchant}, P., {Langer}, N., {Podsiadlowski}, P., {Tauris}, T.~M., \&
  {Moriya}, T.~J. 2016, \aap, 588, A50, \dodoi{10.1051/0004-6361/201628133}

\bibitem[{{M{\'e}sz{\'a}ros} {et~al.}(2019){M{\'e}sz{\'a}ros}, {Fox}, {Hanna},
  \& {Murase}}]{Meszaros_2019}
{M{\'e}sz{\'a}ros}, P., {Fox}, D.~B., {Hanna}, C., \& {Murase}, K. 2019, Nature
  Reviews Physics, 1, 585, \dodoi{10.1038/s42254-019-0101-z}

\bibitem[{{Mikkola} \& {Merritt}(2008)}]{Mikkola_2008}
{Mikkola}, S., \& {Merritt}, D. 2008, \aj, 135, 2398,
  \dodoi{10.1088/0004-6256/135/6/2398}

\bibitem[{Mikkola \& Tanikawa(1999)}]{Mikkola_1999}
Mikkola, S., \& Tanikawa, K. 1999, Monthly Notices of the Royal Astronomical
  Society, 310, 745, \dodoi{10.1046/j.1365-8711.1999.02982.x}

\bibitem[{{Moe} \& {Di Stefano}(2017)}]{Moe_2017}
{Moe}, M., \& {Di Stefano}, R. 2017, \apjs, 230, 15,
  \dodoi{10.3847/1538-4365/aa6fb6}

\bibitem[{{Mondal} {et~al.}(2020){Mondal}, {Belczy{\'n}ski}, {Wiktorowicz},
  {Lasota}, \& {King}}]{Mondal_2020}
{Mondal}, S., {Belczy{\'n}ski}, K., {Wiktorowicz}, G., {Lasota}, J.-P., \&
  {King}, A.~R. 2020, \mnras, 491, 2747, \dodoi{10.1093/mnras/stz3227}

\bibitem[{{Nelson} \& {Eggleton}(2001)}]{Nelson_2001}
{Nelson}, C.~A., \& {Eggleton}, P.~P. 2001, ApJ, 552, 664,
  \dodoi{10.1086/320560}

\bibitem[{{Olejak} \& {Belczynski}(2021)}]{Olejak_2021}
{Olejak}, A., \& {Belczynski}, K. 2021, \apjl, 921, L2,
  \dodoi{10.3847/2041-8213/ac2f48}

\bibitem[{{Olejak} {et~al.}(2021){Olejak}, {Belczynski}, \&
  {Ivanova}}]{Olejak_2021a}
{Olejak}, A., {Belczynski}, K., \& {Ivanova}, N. 2021, \aap, 651, A100,
  \dodoi{10.1051/0004-6361/202140520}

\bibitem[{{Olejak} {et~al.}(2022){Olejak}, {Fryer}, {Belczynski}, \&
  {Baibhav}}]{Olejak_2022}
{Olejak}, A., {Fryer}, C.~L., {Belczynski}, K., \& {Baibhav}, V. 2022, \mnras,
  516, 2252, \dodoi{10.1093/mnras/stac2359}

\bibitem[{{Pavlovskii} {et~al.}(2017){Pavlovskii}, {Ivanova}, {Belczynski}, \&
  {Van}}]{Pavlovskii_2017}
{Pavlovskii}, K., {Ivanova}, N., {Belczynski}, K., \& {Van}, K.~X. 2017,
  \mnras, 465, 2092, \dodoi{10.1093/mnras/stw2786}

\bibitem[{{Paxton} {et~al.}(2015){Paxton}, {Marchant}, {Schwab}, {Bauer},
  {Bildsten}, {Cantiello}, {Dessart}, {Farmer}, {Hu}, {Langer}, {Townsend},
  {Townsley}, \& {Timmes}}]{Paxton_2015}
{Paxton}, B., {Marchant}, P., {Schwab}, J., {et~al.} 2015, \apjs, 220, 15,
  \dodoi{10.1088/0067-0049/220/1/15}

\bibitem[{{Portegies Zwart} {et~al.}(2010){Portegies Zwart}, {McMillan}, \&
  {Gieles}}]{PortegiesZwart_2010}
{Portegies Zwart}, S.~F., {McMillan}, S.~L.~W., \& {Gieles}, M. 2010, \araa,
  48, 431, \dodoi{10.1146/annurev-astro-081309-130834}

\bibitem[{{Rastello} {et~al.}(2021){Rastello}, {Mapelli}, {Di Carlo}, {Iorio},
  {Ballone}, {Giacobbo}, {Santoliquido}, \& {Torniamenti}}]{Rastello_2021}
{Rastello}, S., {Mapelli}, M., {Di Carlo}, U.~N., {et~al.} 2021, \mnras, 507,
  3612, \dodoi{10.1093/mnras/stab2355}

\bibitem[{{Riley} {et~al.}(2021){Riley}, {Mandel}, {Marchant}, {Butler},
  {Nathaniel}, {Neijssel}, {Shortt}, \& {Vigna-G{\'o}mez}}]{Riley_2021}
{Riley}, J., {Mandel}, I., {Marchant}, P., {et~al.} 2021, MNRAS, 505, 663,
  \dodoi{10.1093/mnras/stab1291}

\bibitem[{Rodriguez {et~al.}(2018)Rodriguez, Amaro-Seoane, Chatterjee, \&
  Rasio}]{Rodriguez_2018}
Rodriguez, C.~L., Amaro-Seoane, P., Chatterjee, S., \& Rasio, F.~A. 2018, Phys.
  Rev. Lett., 120, 151101, \dodoi{10.1103/PhysRevLett.120.151101}

\bibitem[{{Roulet} {et~al.}(2021){Roulet}, {Chia}, {Olsen}, {Dai},
  {Venumadhav}, {Zackay}, \& {Zaldarriaga}}]{Roulet2021}
{Roulet}, J., {Chia}, H.~S., {Olsen}, S., {et~al.} 2021, \prd, 104, 083010,
  \dodoi{10.1103/PhysRevD.104.083010}

\bibitem[{{Rozner} {et~al.}(2022){Rozner}, {Generozov}, \&
  {Perets}}]{Rozner_2022}
{Rozner}, M., {Generozov}, A., \& {Perets}, H.~B. 2022, arXiv e-prints,
  arXiv:2212.00807.
\newblock \doarXiv{2212.00807}

\bibitem[{{Sana} \& {Evans}(2011)}]{Sana_2011}
{Sana}, H., \& {Evans}, C.~J. 2011, in IAU Symposium, Vol. 272, Active OB
  Stars: Structure, Evolution, Mass Loss, and Critical Limits, ed. C.~{Neiner},
  G.~{Wade}, G.~{Meynet}, \& G.~{Peters}, 474--485,
  \dodoi{10.1017/S1743921311011124}

\bibitem[{{Sander} {et~al.}(2022){Sander}, {Vink}, {Higgins}, {Shenar},
  {Hamann}, \& {Todt}}]{Sander_2022}
{Sander}, A. A.~C., {Vink}, J.~S., {Higgins}, E.~R., {et~al.} 2022, arXiv
  e-prints, arXiv:2202.04671.
\newblock \doarXiv{2202.04671}

\bibitem[{{Santoliquido} {et~al.}(2021){Santoliquido}, {Mapelli}, {Giacobbo},
  {Bouffanais}, \& {Artale}}]{Santoliquido2021}
{Santoliquido}, F., {Mapelli}, M., {Giacobbo}, N., {Bouffanais}, Y., \&
  {Artale}, M.~C. 2021, \mnras, 502, 4877, \dodoi{10.1093/mnras/stab280}

\bibitem[{{Shao}(2022)}]{Shao_2022}
{Shao}, Y. 2022, Research in Astronomy and Astrophysics, 22, 122002,
  \dodoi{10.1088/1674-4527/ac995e}

\bibitem[{{Spera} {et~al.}(2022){Spera}, {Trani}, \& {Mencagli}}]{Spera_2022}
{Spera}, M., {Trani}, A.~A., \& {Mencagli}, M. 2022, Galaxies, 10, 76,
  \dodoi{10.3390/galaxies10040076}

\bibitem[{{Spruit}(2002)}]{Spruit_2002}
{Spruit}, H.~C. 2002, \aap, 381, 923, \dodoi{10.1051/0004-6361:20011465}

\bibitem[{{Sukhbold} {et~al.}(2016){Sukhbold}, {Ertl}, {Woosley}, {Brown}, \&
  {Janka}}]{Sukhbold_2016}
{Sukhbold}, T., {Ertl}, T., {Woosley}, S.~E., {Brown}, J.~M., \& {Janka}, H.~T.
  2016, \apj, 821, 38, \dodoi{10.3847/0004-637X/821/1/38}

\bibitem[{{Tauris}(2022)}]{Tauris_2022}
{Tauris}, T.~M. 2022, \apj, 938, 66, \dodoi{10.3847/1538-4357/ac86c8}

\bibitem[{{Trani} {et~al.}(2021){Trani}, {Tanikawa}, {Fujii}, {Leigh}, \&
  {Kumamoto}}]{Trani_2021}
{Trani}, A.~A., {Tanikawa}, A., {Fujii}, M.~S., {Leigh}, N.~W.~C., \&
  {Kumamoto}, J. 2021, \mnras, 504, 910, \dodoi{10.1093/mnras/stab967}

\bibitem[{{van Son} {et~al.}(2020){van Son}, {De Mink}, {Broekgaarden},
  {Renzo}, {Justham}, {Laplace}, {Mor{\'a}n-Fraile}, {Hendriks}, \&
  {Farmer}}]{vonSon_2020}
{van Son}, L.~A.~C., {De Mink}, S.~E., {Broekgaarden}, F.~S., {et~al.} 2020,
  \apj, 897, 100, \dodoi{10.3847/1538-4357/ab9809}

\bibitem[{{van Son} {et~al.}(2022){van Son}, {de Mink}, {Callister}, {Justham},
  {Renzo}, {Wagg}, {Broekgaarden}, {Kummer}, {Pakmor}, \&
  {Mandel}}]{vanSon_2022}
{van Son}, L.~A.~C., {de Mink}, S.~E., {Callister}, T., {et~al.} 2022, \apj,
  931, 17, \dodoi{10.3847/1538-4357/ac64a3}

\bibitem[{Vink {et~al.}(2001)Vink, de~Koter, \& Lamers}]{Vink_2001}
Vink, J.~S., de~Koter, A., \& Lamers, H. J. G. L.~M. 2001, Astronomy and
  Astrophysics, 369, 574, \dodoi{10.1051/0004-6361:20010127}

\bibitem[{{Wang} {et~al.}(2021){Wang}, {McKernan}, {Ford}, {Perna}, {Leigh}, \&
  {Mac Low}}]{WangYH_2021}
{Wang}, Y.-H., {McKernan}, B., {Ford}, S., {et~al.} 2021, \apjl, 923, L23,
  \dodoi{10.3847/2041-8213/ac400a}

\bibitem[{{Webbink}(1984)}]{Webbink_1984}
{Webbink}, R.~F. 1984, \apj, 277, 355, \dodoi{10.1086/161701}

\bibitem[{{Xu} \& {Li}(2010)}]{Xu_2010}
{Xu}, X.-J., \& {Li}, X.-D. 2010, \apj, 716, 114,
  \dodoi{10.1088/0004-637X/716/1/114}

\bibitem[{{Zevin} \& {Bavera}(2022)}]{Zevin_2022}
{Zevin}, M., \& {Bavera}, S.~S. 2022, \apj, 933, 86,
  \dodoi{10.3847/1538-4357/ac6f5d}

\end{thebibliography}
\bibliographystyle{aasjournal}

\end{document}